\documentclass[prx,twocolumn,superscriptaddress]{revtex4-2}

\usepackage{tcolorbox}
\usepackage{graphicx}
\usepackage{mathtools}
\usepackage{amsmath}
\usepackage{amssymb}
\usepackage{xcolor}
\usepackage{multirow}

 \usepackage{booktabs}
\usepackage{braket}

\usepackage[pdfencoding=auto, psdextra]{hyperref}
\hypersetup{
	colorlinks=true, 
	linktoc=all,     
	linkcolor=blue,  
}

\newcommand{\uk}{_{\mathbf{k}}}
\newcommand{\uki}{_{\mathbf{k}i}}
\newcommand{\rme}{\text{e}}
\newcommand{\proj}[2]{\ket{#1}\bra{#2}}
\newcommand{\inta}[1]{\int_0^{\infty}\mathrm{d}#1\ }
\DeclareMathAlphabet{\pazocal}{OMS}{zplm}{m}{n}
\newcommand{\ms}{\pazocal}
\newcommand{\mc}{\mathcal}
\newcommand{\mb}{\mathbf}

\begin{document}

\title{Influence of strong molecular vibrations on  decoherence of molecular polaritons}

\author{Dominic M. Rouse}
\email{dominic.rouse@glasgow.ac.uk}
\affiliation{School of Physics and Astronomy, University of Glasgow, Glasgow, G12 8QQ UK}

\author{Erik M. Gauger}
\affiliation{SUPA, Institute of Photonics and Quantum Sciences, Heriot-Watt University, Edinburgh, EH14 4AS, UK}

\author{Brendon W. Lovett}
\email{bwl4@st-andrews.ac.uk}
\affiliation{SUPA, School of Physics and Astronomy, University of St Andrews, St Andrews, KY16 9SS, UK}
\begin{abstract}
We derive the transition rates, dephasing rates, and Lamb shifts for a system consisting of many molecules collectively coupled to a resonant cavity mode. Using a variational polaron master equation, we show that strong vibrational interactions inherent to molecules give rise to multi phonon processes and suppress the light--matter coupling. In the strong light--matter coupling limit, multi-phonon contributions to the transition and dephasing rates strongly dominate over single phonon contributions for typical molecular parameters. This leads to novel dependencies of the rates and spectral line widths on the number of molecules in the cavity. We also find that vibrational Lamb shifts can substantially modify the polariton energies in the strong light--matter coupling limit.
\end{abstract}
\maketitle
\section{Introduction}
%




The confinement of a mode in a microcavity enhances the interaction strength between the light and charged matter within the cavity~\cite{mahan2013many}. The effective interaction strength also scales as the square root of the number of matter systems coupled to the light due to a collective enhancement~\cite{garraway2011dicke}. When the effective interaction strength exceeds dissipation mechanisms and disorder in the joint cavity--matter system, the cavity photons hybridize both in and out of phase with a symmetric superposition of the matter states---the so-called bright state---to form two polariton states. Polaritons inherit properties from both constituents, for example, a small effective mass from the photonic component, whilst retaining the material ability to interact with other polaritons~\cite{kavokin2003cavity}. Additionally, since the bright state is delocalised across many matter systems---often tens-to-thousands of billions---polaritons also afford long range control of matter properties~\cite{rozenman2018long}. The remaining matter state superpositions---known as the dark states---remain uncoupled from the light mode.

Originally demonstrated in an atomic system~\cite{thompson1992observation}, research has recently been directed towards using polaritonic physics to modify and control the properties of molecular systems. For example, superabsorption \cite{quach2020organic}, energy transfer \cite{georgiou2018control, coles2014polariton}, chemical reactivities~\cite{dunkelberger2022vibration,feist2018polaritonic,hertzog2019strong,garcia2021manipulating, fregoni2022theoretical,ebbesen2023introduction}, photophysical dynamics~\cite{thomas2019tilting}, and photoluminescence \cite{herrera2017absorption}. Molecular eigenstates are hybridizations of electronic excitations (excitons) with vibrational states that interact, often strongly, through displacement interactions. The joint excitonic--vibrational eigenstates are called polarons~\cite{kok2010introduction}. 

The current state-of-the-art in analytical theoretical modelling of molecular polaritons is within the assumption of weak vibrational coupling \cite{de2018cavity,martinez2018comment}, or for strong vibrational coupling but a single matter system \cite{denning2020electron}. There has also been numerical calculations for an infinite number of matter systems using mean-field theory \cite{fowler2022efficient}, and to obtain steady state properties and decay rates using the transfer tensor method \cite{wu2024extracting}. 

Weak vibrational couplings induce single phonon transitions between all eigenstates of the joint cavity--molecule system, with the dark states acting as population traps when there are a large number of molecules \cite{de2018cavity,martinez2018comment}. When the effective light--matter interaction strength exceeds the high frequency cutoff of the vibrational baths---a regime now reachable in experiments \cite{shalabney2015coherent}---transitions between eigenstates are strongly suppressed. Refs.~\cite{de2018cavity,martinez2018comment} identify a zero frequency, single phonon contribution to the polariton--ground state decoherence which is suppressed only inversely with the number of molecules. Consequently, this contribution dominates the polariton line widths in spectra when the effective light--matter coupling is strong.

On the other hand, when the vibrational couplings are strong, the effective light--matter couplings between the polarons and light mode are suppressed, and transitions between the eigenstates may also occur through multi phonon processes \cite{denning2020electron}. In practice, strong light--matter coupling in molecular cavity experiments is achieved with $N\sim [10^{10},10^{12}]$ molecules \cite{quach2020organic,shalabney2015coherent} resulting in two polariton states and $N-1\gg 2$ dark states. The role of dark states in the strong vibrational coupling regime remains an open question because for the $N=1$ system in Ref.~\cite{denning2020electron} there are no dark states.

\begin{figure}[ht!]\centering
	\includegraphics[width=\columnwidth]{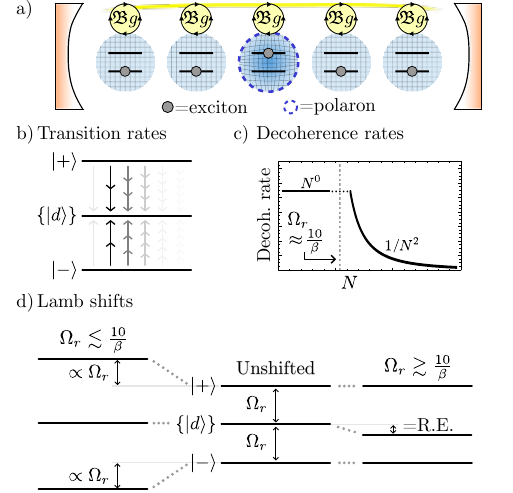}
	\caption{An illustration of our model and main results. (a) Our model consists of $N$ two-level systems, each coupled to a common cavity mode which causes polariton formation, and to a local vibrational bath which causes polaron formation. The phonon dressing within the polaron suppresses the light--matter coupling strength $g$ by a factor $0<\mathfrak{B}< 1$. In the large $N$ limit our main results are as follows. (b) The $N-1$ degenerate dark states $\{\ket{d}\}$ act as population traps (in agreement with Refs.~\cite{de2018cavity,martinez2018comment}) but, if $\Omega_r=g\mathfrak{B}\sqrt{N}\gg\omega_0$ where $\omega_0$ is the high frequency cutoff of the vibrational baths, then multi phonon processes strongly dominate over single phonon processes. Typically processes with $2$ or $3$ phonons dominate, indicated by the saturation of the arrow colors. (c) The decoherence rates have different $N$ scaling depending on the size of the collective light--matter coupling $\Omega_r$ compared to the inverse temperature $\beta$. When $\Omega_r\lesssim 10/\beta$, decoherence is dominated by processes with even numbers of phonons, whilst when $\Omega_r\gtrsim 10/\beta$, \textit{only} two-phonon processes contribute significantly. When $\Omega_r\sim\Omega_\beta$ the $N$ dependence is more complicated and not illustrated. (d) When $\Omega_r\lesssim 10/\beta$, the polariton states are Lamb shifted by an equal and opposite amount proportional to the bare splitting $\Omega_r$ which can be substantial for moderately strong vibrational coupling, whilst when $\Omega_r\gtrsim10/\beta$, only the dark states are shifted by the vibrational reorganization energy (`R.E.'). All Lamb shifts shown result from transitions involving the dark states. Typical molecular experiments operate within the regime $\Omega_r\lesssim 10/\beta$ \cite{quach2020organic,shalabney2015coherent} but can reach $\Omega_r\gtrsim\omega_0$ \cite{shalabney2015coherent}.} \label{fig:sketch}
\end{figure}

In this paper we derive analytical expressions for the key dynamical rates and Lamb shifts that remain accurate when the vibrational coupling is strong and for an arbitrary number of molecules in the cavity. We analyze our expressions for the parameter regimes most relevant to recent experiments \cite{quach2020organic,shalabney2015coherent}. Our methodology builds upon Refs.~\cite{de2018cavity,martinez2018comment} by utilizing a variational polaron transformation, and upon Ref.~\cite{denning2020electron} by including an arbitrary number of molecules. Our model and main results are illustrated in Fig.~\ref{fig:sketch}. 

This paper is organized as follows. In Section~\ref{sec:model} we introduce the Hamiltonian of the model, and in Section~\ref{sec:weak} we summarize Refs.~\cite{de2018cavity,martinez2018comment} by deriving the transition rates, decoherence rates, and Lamb shifts valid when the vibrational couplings are weak. In Section~\ref{sec:VPT} we transform the Hamiltonian into the variational polaron frame and identify parameter regimes with distinct dependencies on the light--matter and vibrational couplings. We also show that when the light--matter and vibrational couplings are simultaneously strong, ``resonant'' cavity experiments should be modelled by non-resonant Hamiltonians. In Section~\ref{sec:VPME} we study the expressions for transition rates, decoherence rates, and Lamb shifts in the variational polaron master equation, with an emphasis on presenting simple and generic conclusions for the parameter regime most relevant to experiments. In Section~\ref{sec:NonResonance} we discuss corrections for non-resonant systems. Finally we summarize our results in Section~\ref{sec:conclusion}.

\section{Model}\label{sec:model}
We consider $N$ identical molecules treated as two-level systems with transition energy $\omega_m$ and independent but identical vibrational baths, see Fig.~\ref{fig:sketch}(a). The transition dipoles of the molecules couple to a cavity mode with energy $\omega_c$ with a light--matter coupling of $g$. We neglect permanent dipole interactions with the cavity which is a typical assumption within molecular polaritonics \cite{fregoni2022theoretical}. This is the localized bath model considered in Refs.~\cite{de2018cavity,martinez2018comment}. 

We partition the Hamiltonian as $H=H_S+H_B+H_{SB}$. The system, which consists of the quantised cavity mode and molecules, is described by
\begin{equation}\label{eq:HS}
	H_S=\omega_c a^\dagger a+\sum_{i=1}^N\left[\omega_m\sigma^+_i\sigma_i^-+g\left(a\sigma^+_i+a^\dagger\sigma^-_i\right)\right],
\end{equation}
where $a^\dagger$ and $\sigma^+_i$ create an excitation in the cavity mode and $i\text{th}$ molecule, respectively. Later we will enforce resonance between the cavity mode and molecular transition. However, the requirement for resonance differs for weak and strong vibrational couplings, and so we discuss each in the appropriate sections. 

The vibrational baths of the molecules are described by
\begin{equation}\label{eq:HB}
	H_{B}=\sum_{i=1}^N\sum\uk \omega\uk b\uki^\dagger b\uki,
\end{equation}
where $b^\dagger\uki$ creates a phonon of wavevector $\mb{k}$ and energy $\omega\uk$ in the vibrational bath of the $i$th molecule. The displacement interactions induced by the vibrational baths are described by
\begin{equation}\label{eq:HSB}
	H_{SB}=\sum_{i=1}^N\sigma^+_i\sigma^-_i\sum\uk f\uk\left(b\uki^\dagger+b\uki\right),
\end{equation}
where $f\uk$ is the coupling strength of a mode with wavevector $\mb{k}$. The vibrational coupling of each molecule to its local bath is characterised by the spectral density, $J(\omega)=\sum\uk f\uk^2\delta(\omega-\omega\uk)$. We consider spectral densities of the form
\begin{equation}\label{eq:J}
	J(\omega)=A\Theta(\omega)\frac{\omega^p}{\omega^{p-1}_0}\rme^{-\frac{\omega^2}{\omega_0^2}},
\end{equation}
where $\omega_0$ is the high frequency cut-off, $p$ is the Ohmicity, $\Theta(\omega)$ is the Heaviside step function, and $A$ is a dimensionless coupling constant with $A\gtrsim 0.1$ signalling strong coupling. We choose Eq.~\eqref{eq:J} to make connection with Refs.~\cite{de2018cavity,martinez2018comment} where the Ohmic ($p=1$) case is studied at weak vibrational coupling, although our results hold for any spectral density which does not suffer from the well-known infrared divergence of the variational polaron transformation \cite{nazir2012ground,silbey1989tunneling}. Eq.~\eqref{eq:J} is suitable for modelling the broad low-frequency contribution to molecular spectral densities \cite{renger2002relation}, but omits the peaked structure that can generate strongly non-Markovian dynamics \cite{lorenzoni2024systematic,ratsep2007electron,sowa2018beyond}. 

Realistic values for the model parameters can be identified from recent experiments \cite{quach2020organic,shalabney2015coherent}. In Refs.~\cite{quach2020organic} and \cite{shalabney2015coherent}, phonon renormalized light--matter couplings of $g\mathfrak{B}=0.01~\mu\text{eV}$ and $g\mathfrak{B}=0.1~\mu\text{eV}$, and maximum collective light--matter coupling strengths ($\Omega_r=g\mathfrak{B}\sqrt{N}$) of $4.2~\text{meV}$ and $20.7~\text{meV}$ were found, respectively. As we demonstrate in Appendix~\ref{app:QB}, we estimate $A=0.083$ for the dye molecules in Ref.~\cite{quach2020organic} which is near to the strong vibrational coupling regime. In Ref.~\cite{shalabney2015coherent}, the cut-off frequency for the bath is $\omega_0=6~\text{meV}$, and the experiments in both Refs.~\cite{quach2020organic} and \cite{shalabney2015coherent} were performed at room temperature $T=1/\beta=0.0258~\text{eV}$ which is typical in molecular polaritonics. We will frequently refer to \textit{typical molecular parameters} which we take as the following: bare light--matter coupling $g=0.1~\mu\text{eV}$, high frequency cutoff $\omega_0=6~\text{meV}$, vibrational coupling strength $A=0.083$, and temperature $T=300~\text{K}$.


\section{The Weak vibrational coupling master equation}\label{sec:weak}
In this section we summarize the results of Refs.~\cite{de2018cavity,martinez2018comment} relevant to our study by deriving the Redfield master equation resulting from $H_{SB}$ in Eq.~\eqref{eq:HSB} perturbing $H_S+H_B$ in Eqs.~\eqref{eq:HS}--\eqref{eq:HB}. This weak vibrational coupling master equation (WCME) becomes inaccurate once the vibrational and collective light--matter couplings become comparable. We will introduce a parameter to quantify this comparison in Section~\ref{sec:VPT}.

\begin{figure}[ht!]\centering
	\includegraphics[width=0.45\textwidth]{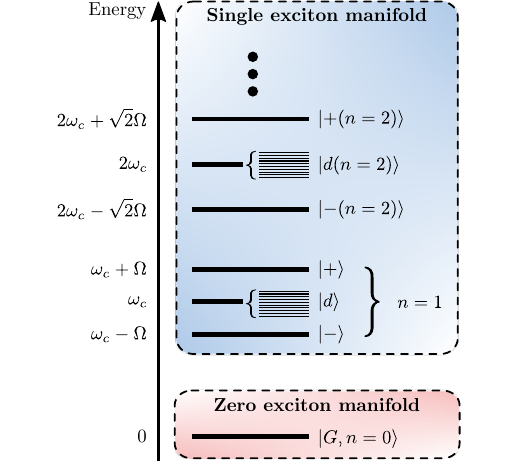}
	\caption{The zero and single exciton manifolds of $H_S$ when the cavity mode is resonant with the molecular transition energy. Due to the symmetry of $H_S$ each set of $N+1$ states identified by photon number $n\ge 1$ are decoupled. To study the linear response of the system we study the dynamics for $n=1$. } \label{fig:eigensystem}
\end{figure}

Deriving the master equation requires diagonalizing $H_S$ in Eq.~\eqref{eq:HS}. We assume that the total number of excitons at any given time does not exceed one \cite{de2018cavity}, which, because $H_S$ preserves the total number of excitations, decouples the eigenstates into sets uniquely identified by photon number $n$. Each set is spanned by $N+1$ states, $\{\ket{G,n},\ket{e_i,n-1}_{i=1,\ldots,N}\}$, where $\ket{\text{mol},n}=\ket{\text{mol}}\otimes\ket{n}$ is a product state with $n$ photons and either zero excitons ($\ket{\text{mol}}=\ket{G}$) or an exciton in the $i$th molecular state only ($\ket{\text{mol}}=\ket{e_i}$). The set relevant to the master equation is determined by the number of photons in the cavity, which we assume to be constant on timescales induced by the vibrational interactions. Following Ref.~\cite{de2018cavity} we choose $n=1$, but, as shown in Fig.~\ref{fig:eigensystem}, the states with $n>1$ differ only by constant factors in the transition energies \footnote{The transition energies between the eigenstates increases when there are more photons in the cavity. This will change whether single phonon or multi phonon processes dominate the transition rates, which, as we later show, depends on the sizes of the transition energies compared to $\omega_0$.}. 

The $n=1$ single excitation subspace of $H_S$ contains an upper polariton $\ket{+}$, lower polariton $\ket{-}$, and $N-1$ degenerate dark states $\{\ket{d}\}$ for $d\in\{d_1\ldots d_{N-1}\}$. The upper and lower polaritons are symmetric and antisymmetric superpositions of single excitation states,
\begin{equation}
\ket{\pm}=\frac{1}{\sqrt{2}}\left(\ket{G,1}\pm\ket{B}\right),
\end{equation} 
where $\ket{B}=\sum_{i=1}^N\ket{e_i,0}/\sqrt{N}$ is the bright state. The dark states,
\begin{equation}
	 \ket{d}=\sum_{i=1}^Nu_{id}\ket{e_i,0},
	\end{equation}
	are the $N-1$ degenerate superpositions of single exciton states orthonormal to $\ket{B}$.  The coefficients $u_{id}$ are complex valued and satisfy
\begin{align}
	\sum_{d}u_{id}u^*_{jd}&=\begin{cases}
		-\frac{1}{N} &\text{ if }i\neq j\\
		\frac{N-1}{N} &\text{ if }i=j,
	\end{cases}\label{eq:i1}\\
	\sum_{i=1}^Nu_{id}u^*_{id'}&=\delta_{dd'},\label{eq:i2}\\
	\sum_iu_{id}&=0,\label{eq:i3}
\end{align}
which enforces unit trace of the density operator (Eq.~\eqref{eq:i1}) and orthonormality of the eigenstates (Eqs.~\eqref{eq:i2}--\eqref{eq:i3}). 

Resonance between the cavity mode and molecular transitions in Eq.~\eqref{eq:HS} is enforced by choosing $\omega_c=\omega_m$. On resonance, the polariton energies are $\omega_\pm=\omega_c\pm\Omega$ where
\begin{equation}\label{eq:RabiW}
\Omega=g\sqrt{N}, 
\end{equation}is the collective light--matter coupling, and, regardless of resonance, the dark states have an energy $\omega_d=\omega_m$.

Denoting the reduced density operator for the light--matter system as $\rho_S(t)=\text{Tr}_B[\rho(t)]$---where $\text{Tr}_B[\cdot]$ is the trace over the joint Hilbert spaces of the vibrational baths---one finds that the WCME in the Schr\"odinger picture \cite{breuer2002theory} is
\begin{multline}\label{eq:MEweak}
\dot{\rho}_S(t)=-i\left[H_S,\rho_S(t)\right]\\+\sum_{\alpha,\beta,\gamma,\delta}c_{\alpha\beta\gamma\delta}\Gamma_1(\omega_\delta-\omega_\gamma)\left[\Pi_{\gamma\delta}\varrho_S(t),\Pi_{\alpha\beta}\right]+\text{H.c.},
\end{multline}
where $\Pi_{\alpha\beta}=\proj{\alpha}{\beta}$ is an eigenstate transition operator, $\omega_\delta$ is the energy of eigenstate $\ket{\delta}$, Greek indices sum over all eigenstates in the $n=1$ single exciton manifold, and `H.c.' denotes the Hermitian conjugate. The coefficients are
\begin{equation}\label{eq:c}
	c_{\alpha\beta\gamma\delta}=\sum_{i=1}^Nu_{i\alpha}u_{i\beta}^*u_{i\gamma}u_{i\delta}^*,
\end{equation}
where $u_{i\alpha}=\braket{e_i,0|\alpha}$ such that $u_{i\pm}=\pm 1/ \sqrt{2N}$ and $u_{id}$ satisfy Eqs.~\eqref{eq:i1}--\eqref{eq:i3}. The Fourier transform of the bath correlation function is
\begin{equation}\label{eq:G1}
\Gamma_1(\nu)=\frac{1}{2}\gamma_1(\nu)+iS_1(\nu),
\end{equation}
where
\begin{equation}\label{eq:g1}
\gamma_1(\nu)=2\pi \times\begin{cases}
J(\nu)\tilde{n}_B(\nu)]&\text{if }\nu\ge0,\\
J(|\nu|)n_B(|\nu|)&\text{if }\nu<0,
\end{cases}
\end{equation}
with $n_B(\nu)= 1/(\exp(\beta\nu)-1)$ the Bose-Einstein distribution, $\tilde{n}_B(\nu)=1+n_B(\nu)$, and
\begin{equation}\label{eq:S}
	S_1(\nu)=\mathcal{P}\inta{\omega}J(\omega)\left[\frac{\tilde{n}_B(\omega)}{\nu-\omega}+\frac{n_B(\omega)}{\nu+\omega}\right],
\end{equation}
where $\mathcal{P}$ denotes the principal value.

The subscript `$1$' denotes that the correlation function originates from single phonon interactions. Eq.~\eqref{eq:MEweak} is the same non-secular master equation as derived in Refs.~\cite{de2018cavity,martinez2018comment}. In this paper, we are only interested in the transition rates, dephasing rates, and Lamb shifts of the eigenstates, which are secular contributions to the master equation. These terms can be obtained from any master equation by deriving the coefficient of the element $\rho_{\mu\nu}(t)\equiv\braket{\mu|\rho_S(t)|\nu}$ within the equation of motion for $\dot{\rho}_{\mu\nu}(t)$, where $\mu$ and $\nu$ label any eigenstates of $H_S$. This term takes the general form
\begin{equation}\label{eq:dotrho}
\dot{\rho}_{\mu\nu}(t)= - r_{\mu\nu} \rho_{\mu\nu}(t)+\ldots,
\end{equation}
where
\begin{equation}\label{eq:Gmunu}
	r_{\mu\nu}=\frac{k_\mu^\downarrow+k_{\nu}^\downarrow}{2}+k^\phi_{\mu\nu}+i\delta_{\mu\nu},
\end{equation}
and terms in the ellipsis in \eqref{eq:dotrho} do not depend on $\rho_{\mu\nu}(t)$. In Eq.~\eqref{eq:Gmunu} we have defined the total loss rate of state $\ket{\mu}$ as,
\begin{equation}
k_\mu^\downarrow=\sum_{\alpha\neq\mu}k_{\mu\to\alpha},
\end{equation} 
where $k_{\mu\to\alpha}$ is the transition rate from state $\ket{\mu}$ to $\ket{\alpha}$. We have also defined the dephasing rate, $k_{\mu\nu}^\phi$, of the coherence between states $\ket{\mu}$ and $\ket{\nu}$ with the properties $k_{\mu\mu}^\phi=0$ and $k_{\mu\nu}^\phi=k_{\nu\mu}^\phi$. The last term in Eq.~\eqref{eq:Gmunu} is the Lamb shifted transition energy from state $\ket{\mu}$ to $\ket{\nu}$, given by $\delta_{\mu\nu}=(\omega_\mu+\lambda_\mu)-(\omega_\nu+\lambda_\nu)$ where $\lambda_\mu$ is the Lamb shift of state $\ket{\mu}$. In the following subsections we analyse the expressions for the quantities appearing in Eq.~\eqref{eq:Gmunu}.

\subsection{Transition rates}
The transition rates obtained from the WCME in Eq.~\eqref{eq:MEweak} are
\begin{equation}
	k_{\mu\to\alpha}=c_{\mu\alpha\alpha\mu}\gamma_1(\omega_\mu-\omega_\alpha)\label{eq:transW}.
\end{equation}
Evaluating $c_{\mu\alpha\alpha\mu}$ using Eq.~\eqref{eq:c} we find
\begin{align}
k_{\pm\to\mp}&=\frac{1}{4N}\gamma_1(\pm2\Omega),\label{eq:gw1}\\
k_{\pm\to d}&=k_{d\to\mp}=\frac{1}{2N}\gamma_1(\pm\Omega),\label{eq:gw2}\\
	k_{d\to d'\neq d}&=\frac{1}{N}\gamma_1(0),\label{eq:gw4}
\end{align} 
and all transition rates involving the zero excitation state $\ket{G,0}$ are zero. Eqs.~\eqref{eq:gw1}--\eqref{eq:gw2} describe decay by single phonon emission (positive frequency arguments) and excitation by single phonon absorption (negative frequency arguments). These expressions take the form of Fermi's Golden Rule. Eq.~\eqref{eq:gw4} describes transitions between degenerate states, which, being zero frequency transitions, do not contribute to overall population transfer. However, these transitions will contribute to decoherence and Lamb shifts. 

From Eqs.~\eqref{eq:gw1}--\eqref{eq:gw4} we obtain the following loss rates for each state
\begin{align}
	k_\pm^\downarrow&=\frac{1}{4N}\gamma_1(\pm 2\Omega)+\frac{N-1}{2N}\gamma_1(\pm\Omega),\label{eq:kpmw}\\
	k_d^\downarrow&=\frac{1}{2N}\left[\gamma_1(\Omega)+\gamma_1(-\Omega)\right]+\frac{N-2}{N}\gamma_1(0).\label{eq:kdw}
\end{align}
As noted by Ref.~\cite{de2018cavity}, since there are $(N-1)/2$ times more dark states than polaritons, in the large $N$ limit the dark states act as population traps. 

\subsection{Dephasing rates}
The dephasing rates obtained from the WCME are
\begin{equation}\label{eq:gWPhi}
k_{\mu\nu}^\phi=\gamma_1(0)
\left[\frac{c_{\mu\mu\mu\mu}+c_{\nu\nu\nu\nu}}{2}-c_{\mu\mu\nu\nu}\right].	
	\end{equation}
We now briefly introduce terminology to distinguish the three possible contributions to decoherence. Generally, the decoherence rate of the coherence between states $\ket{\mu}$ and $\ket{\nu}$ is the real part of $r_{\mu\nu}$ in Eq.~\eqref{eq:Gmunu} for $\mu\neq\nu$. The three possible contributions to this are: (1) transitions from either state, (2) virtual transitions from either state back to itself, or (3) other pure dephasing contributions. Whilst transitions contribute to decoherence they are not dephasing processes. The first term in Eq.~\eqref{eq:gWPhi} arises from virtual transitions from $\ket{\mu}\to\ket{\mu}$ and $\ket{\nu}\to\ket{\nu}$, and the second term from other pure dephasing contributions. Both dephasing processes depend on the zero frequency bath correlation function. 

Using Eq.~\eqref{eq:c} to evaluate $c_{\mu\mu\mu\mu}$ and $c_{\mu\mu\nu\nu}$ we find the following dephasing rates
\begin{align}
	k_{+-}^\phi&=k_{d_id_j}^\phi=0,\label{eq:gdeph1}\\
	k^
	\phi_{\pm G}&=k^\phi_{\pm d}=\frac{1}{8N}\gamma_1(0),\label{eq:gDP}\\
	k_{dG}^\phi&=\frac{1}{2N}\gamma_1(0),\label{eq:gdeph3}
\end{align}
where subscript `$G$' refers to the zero excitation state $\ket{G,0}$.  The dephasing rate of the coherence between $\ket{\pm}$ and $\ket{\mp}$, denoted $k_{+-}^\phi$, is zero because only the molecular parts of the polariton wavefunctions interact with the vibrational baths---see $H_{SB}$ in Eq.~\eqref{eq:HSB}---and both polaritons feature the same molecular wavefunction up to a phase \cite{martinez2018comment}. This symmetry will be broken by the variational polaron transformation. 

As we show in Appendix~\ref{app:spectrum}, within the validity of the WCME---where phonon sidebands are small, and photon leakage and non-radiative molecular decay are slower than vibrational decoherence---and of the quantum regression theorem---where the Born and Markov approximations hold \cite{mccutcheon2016optical}---the absorption spectrum of the cavity is given by
\begin{equation}\label{eq:A2}
\ms{A}(\omega)=\frac{\ms{A}_0}{2}\sum_{p\in\{+,-\}}\frac{\text{Re}[r_{pG}]}{\text{Re}[r_{pG}]^2+\left(\delta_{pG}-\omega\right)^2},
\end{equation}
where $\ms{A}_0$ is the intensity. The spectrum describes two Lorentzian peaks centered at the Lamb shifted polariton energies, with full width half maxima equal to $2\text{Re}[r_{\pm G}]=k_\pm^\downarrow+2k_{\pm G}^\phi$. As discussed in Ref.~\cite{de2018cavity}, Eq.~\eqref{eq:gDP} predicts that when $N$ is large enough to satisfy $\Omega\gg\omega_0$---so that transition rates are exponentially suppressed by the high frequency cutoff of the vibrations---the width of the polariton peaks in the cavity absorption spectrum  will be dominated by $k_{\pm G}^\phi$ in Eq.~\eqref{eq:gDP}, which scales as $1/N$. As illustrated in Fig.~\ref{fig:sketch}(c) this prediction will be challenged by the variational polaron master equation.

The single phonon dephasing rates in Eqs.~\eqref{eq:gDP}--\eqref{eq:gdeph3} depend on
\begin{equation}\label{eq:g10}
\gamma_1(0)=\frac{2\pi}{\beta}\lim_{\omega\to 0}\frac{J(\omega)}{\omega}.
\end{equation}
Eq.~\eqref{eq:g10} is only non-zero and non-divergent if the spectral density is exactly linear in frequency at small frequencies, which is unlikely to be the case in molecular systems. 
Generally, pure dephasing cannot have single-phonon Markovian contributions, because it results from interactions between the system and the baths which do not lead to energy exchange. Therefore, single-phonon Markovian pure dephasing could only occur through emission, or, equivalently absorption, of a zero frequency phonon, resulting in the spurious expression in Eq.~\eqref{eq:g10} which varies discontinuously as the Ohmicity changes. Conversely, single-phonon \textit{non}-Markovian pure dephasing is entirely possible---the system emits a finite energy phonon which is later re-absorbed---and so is \textit{multi}-phonon Markovian pure dephasing---the system simultaneously emits and absorbs an equal number of phonons of the same finite frequency.

Single-phonon non-Markovian processes can be described by the WCME in Eq.~\eqref{eq:MEweak} if the Markovian assumption in the bath correlation function is relaxed. However, multi-phonon Markovian processes require a strong vibrational coupling theory. We defer the discussion of both processes to Section~\ref{sec:VPME} so that the present section remains a faithful summary of Refs.~\cite{de2018cavity,martinez2018comment}.



\subsection{Lamb shifts}\label{sec:LambWeak}
The Lamb shifts obtained from the WCME are 
\begin{equation}\label{eq:lambW}
\lambda_\mu=\sum_\alpha c_{\mu\alpha\alpha\mu}S_1(\Delta_{\mu\alpha}),
\end{equation}
where $S_1(\nu)$ is given in Eq.~\eqref{eq:S}. The frequency dependence of $S_1(\nu)$ is generally complicated, but, when $|\nu|\gg\omega_0$, one can ignore the $\pm\omega$ in the denominators of the integrand in Eq.~\eqref{eq:S}, resulting in $S_1(\nu)\sim 1/\nu$ for any spectral density. Consequently, when $\Omega\gg\omega_0$, Lamb shifts resulting from terms with $\mu\neq \alpha$ in Eq.~\eqref{eq:lambW} scale as $c_{\mu\alpha\alpha\mu}/\Omega$ which will be negligible in comparison to the bare energy splittings, scaling as $\Omega$, when $N$ is large. Additionally, since $u_{i\pm}=\pm1/\sqrt{2 N}$, the only Lamb shifts with $\mu=\alpha$ that may be comparable to the bare energy splitting at large $N$ are the contributions to $\lambda_d$ resulting from real and virtual dark state transitions. One finds that these contributions are
\begin{equation}\label{eq:S10}
	\lambda_d\approx\lim_{N\gg 1}\frac{N-1}{N}S_1(0)\to -\inta{\omega}\frac{J(\omega)}{\omega},
\end{equation}
which is the negative of the reorganisation energy of the vibrational baths. Therefore, if the vibrational coupling is comparable to the Rabi frequency, $|S_1(0)|\gtrsim\Omega$, then the dark state energy may be non-negligibly Lamb shifted. However, in this regime the WCME is no longer accurate.

\subsection{Summary of weak vibrational coupling} 
We have derived the Redfield equation in the limit of weak vibrational coupling and recovered the same expressions for the transition rates, decoherence rates, and Lamb shifts found in Refs.~\cite{de2018cavity,martinez2018comment}. There remain a number of questions unanswered by the weak vibrational coupling theory. 

(1)~For the single matter system model in Ref.~\cite{denning2020electron}, strong vibrational coupling suppressed the bare light--matter coupling strength and introduced multi phonon processes. How does the suppression scale with the number of molecules? And, do multi phonon processes qualitatively change the dynamics? 

(2)~When $\Omega\gg \omega_0$, the WCME predicts that the decoherence rates---and so the line widths of the polaritons---are dominated by the $1/N$ pure dephasing contribution from the single-phonon Markovian process described by $\gamma_1(0)$ given in Eq.~\eqref{eq:g10}. This term produces divergences or zero values for most spectral density types. What happens if the Markovian assumption is relaxed? And, to leading order, is decoherence a multi phonon process? 

(3) Are the Lamb shifts still negligible in the large $N$ limit when the vibrational coupling is strong?

The answer to these questions requires a strong vibrational coupling theory.  

\section{Variational polaron theory}\label{sec:VPT}
A polaron master equation is a Redfield equation derived after transforming the Hamiltonian $H$ by a state-dependent phonon displacement. The resulting master equation is perturbative in a quantity that remains small when the vibrational coupling is strong \cite{nazir2016modelling}. However, the perturbative quantity grows proportional to any driving in the system, which, in our model, is the collective light--matter coupling. This breakdown can be mitigated by employing a variational version of the unitary transformation \cite{pollock2013multi,mccutcheon2011variational,mccutcheon2011general}. The unitary operator is optimised such that the Gibbs state of the unperturbed Hamiltonian is as close to the equilibrium steady state as permitted by a polaron--type transformation, and so the residual interaction is more amenable to perturbation theory. 

The Hamiltonian in the variational frame is
$\mc{H}=U^\dagger H U$ where
\begin{equation}\label{eq:Uprime1} U=\sum_i \left(\sigma^-_i\sigma^+_i+B_i\sigma^+_i\sigma^-_i\right),
\end{equation}
and $B_i=\exp[-\sum\uk(\eta\uk/\omega\uk)(b\uki^\dagger-b\uki)]$ is a displacement operator. We use calligraphic notation to denote operators transformed into the variational frame. Eq.~\eqref{eq:Uprime1} describes a transformation that displaces a vibrational bath when the corresponding molecule is in its excited state, but otherwise does not transform the bath. The variational parameters, $\eta\uk$, are free parameters used to optimize the transformation, and have the general form $\eta\uk = G(\omega\uk)f\uk$ where 
\begin{equation}
	G(\omega)=\frac{ \omega}{ \omega+\bar{G}\coth\left(\frac{\beta\omega}{2}\right)}.\label{eq:OptRes}
\end{equation}
The intuition within Eq.~\eqref{eq:OptRes} is that, after molecular excitation, the low frequency phonon modes displace more slowly than high frequency ones, with the boundary between slow and fast modes determined by the parameter $\bar{G}$, which we will define soon.

After applying the transformation in Eq.~\eqref{eq:Uprime1} we find the polaron Hamiltonian $\mc{H}=\mc{H}_S+\mc{H}_B+\mc{H}_{SB}$. The system part is the Tavis--Cummings Hamiltonian with renormalized molecular energy and light--matter coupling,
\begin{multline}\label{eq:HSp}
	\mc{H}_S=\omega_c a^{\dagger}a+\sum_{i=1}^N\big[\left(\omega_m-\lambda^v\right)\sigma^+_i\sigma^-_i\\+\mathfrak{B}g\left(a\sigma^+_i+a^{\dagger}\sigma^-_i\right)\big],
\end{multline}
where 
\begin{equation}\label{eq:lv}
	\lambda^v=\inta{\omega}\frac{J(\omega)}{\omega}G(\omega)\left(2-G(\omega)\right),
\end{equation} 
and $0< \mathfrak{B}< 1$ is a suppression of the light--matter coupling by the vibrational coupling, given by
\begin{align}
	\mathfrak{B}=\exp\left[-\frac{1}{2}\int_0^\infty\text{d}\omega\ \frac{J(\omega)G^2(\omega)}{\omega^2}\coth\left(\frac{\beta\omega}{2}\right)\right].\label{eq:Bcont}
\end{align} 
The bath part of the Hamiltonian is the same as before the transformation, $\mc{H}_B=H_B$, and the interaction Hamiltonian has two components, $\mc{H}_{SB}=\mc{H}_{D}+\mc{H}_{P}$. There is a displacement--type interaction,
\begin{equation}\label{eq:VE}
	\mc{H}_{D}=\sum_{i=1}^N\sigma^+_i\sigma^-_i\sum_{\mb{k}}\left(f\uk-\eta\uk\right)(b\uki^{\dagger}+b\uki),
\end{equation}
named due to its similarity to Eq.~\eqref{eq:HSB}, and a polaron--type interaction,
\begin{equation}\label{eq:VP}
	\mc{H}_{P}=\sum_{i=1}^Ng\left(a\left[B_i^\dagger-\mathfrak{B}\right]\sigma^+_i+a^{\dagger}\left[B_i-\mathfrak{B}\right]\sigma^-_i\right).
\end{equation}

 When $\eta\uk\to 0$ the polaron--type interaction vanishes ($\mc{H}_{P}\to 0$) and the Hamiltonian reverts back to its original form ($\mc{H}\to H$). Conversely, when $\eta\uk\to f\uk$ the displacement--type interaction vanishes ($\mc{H}_{D}\to0$) as the polaron incorporates the total energy of the displacement described by $H_{SB}$. Generally, the $\eta\uk$ range from $0$ to $f\uk$ as the frequency of the mode increases, such that both interaction types contribute to the dynamics with a weighting determined by $G(\omega)$ \footnote{We have redefined the partition between the system and interaction Hamiltonians, such that $\mc{H}_\alpha\neq U^\dagger H_\alpha U$ for any $\alpha\in\{S,B,SB\}$. This partition is made to ensure $\text{tr}[\mc{H}_{E}\rho_B]=\text{tr}[\mc{H}_{P}\rho_B]=0$---where $\rho_B=\exp(-\beta \mc{H}_B)/\text{tr}[\exp(-\beta \mc{H}_B)]$ and $\text{tr}_B[\cdot]$ is the partial trace over all baths---so that
 perturbation theory in $\mc{H}_{SB}$ yields the familiar Redfield equation.}.
 
 In the vibrational coupling theory in Section~\ref{sec:weak}, resonance in $H_S$ was enforced by choosing the cavity energy $\omega_c=\omega_m$ where $\omega_m$ is the energy of the molecular excitation. Clearly, the same choice for $\omega_c$ in $\ms{H}_S$ in Eq.~\eqref{eq:HSp} does not yield a resonant Hamiltonian. Moreover, since the variational polaron frame molecular energy $\omega_m-\lambda^v$ depends on $G(\omega)$, the value of $\omega_c$ that leads to resonance implicitly depends on itself \footnote{Alternatively, one can build a wedge-shaped cavity where the mode energy can be varied continuously until resonance is achieved. However, since we later find that the Hamiltonian is essentially resonant for parameter regimes relevant to experiments, the Hamiltonians for a resonant wedge-shaped cavity and for a cavity with an energy tuned to the bare molecular energy, are the same.}. This leads one to consider how to correctly enforce resonance. The relevant question is, how, in the experiments we are modelling, is the molecular energy determined? 
 
 To answer this question it is helpful to consider the Hamiltonian for an isolated molecule, \begin{multline}\label{eq:Hmol}
 	H_\text{mol}=\omega_m\sigma_+\sigma_-+\sum_\mb{k}\omega\uk b\uk^\dagger b\uk \\+\sigma_+\sigma_-\sum\uk f\uk(b\uk^\dagger+b\uk).
 	\end{multline}
 $H_\text{mol}$ is diagonalized by a full polaron transformation---the variational polaron transformation with $G(\omega)=1$---yielding $\ms{H}_\text{mol}=(\omega_m+S_1(0))\sigma_+\sigma_-+\sum_\mb{k}\omega\uk b\uk^\dagger b\uk$ where $S_1(0)$ is given in Eq.~\eqref{eq:S10}. In a measurement of the molecular energy---for example, by coupling the molecule to a weak probe field and measuring the absorption spectrum---the vibrational reorganisation energy $S_1(0)$ cannot be separated from the excitonic energy $\omega_m$. Therefore, after assuming that the Lamb shift induced by the weak probe field is negligible, one would attempt to enforce resonance by building a cavity with $\omega_c=\omega_m+S_1(0)$. This is not the resonance condition for $H_S$ in the weak vibrational coupling theory considered in Section~\ref{sec:weak}. However, as we will soon show, if $|S_1(0)|\ll 2\Omega$---a good definition of weak vibrational coupling---then $\omega_c=\omega_m$ is a good approximation to resonance in the weak vibrational coupling regime. 
 
In variational polaron theory, $\omega_c=\omega_m+S_1(0)$ is also not generally the resonance condition for $\ms{H}_S$. Consequently, ``resonant" experiments should be modelled by a non-resonant Hamiltonian in the variational polaron frame. The resulting detuning between the cavity and molecular transition is $\Delta=\omega_m-\lambda^v-\omega_c$, and substituting in $\omega_c=\omega_m+S_1(0)$ yields
\begin{equation}\label{eq:Delta}
	\Delta=\inta{\omega}\frac{J(\omega)}{\omega}\left(1-G(\omega)\right)^2.
\end{equation}
Eq.~\eqref{eq:Delta} shows that the system will be non-resonant when the vibrational and light--matter coupling strengths are strong and comparable because a large $\Delta$ requires a simultaneously small $G(\omega)$ and large $J(\omega)$.

To complete the transformation we must define $\bar{G}$ in Eq.~\eqref{eq:OptRes}. This is found through an optimization scheme detailed in Appendix~\ref{app:VO}, yielding
\begin{equation}\label{eq:Gbar}
	\bar{G}=\frac{\frac{2\Omega_r^2}{\theta}\sinh\left(\frac{\beta\theta}{2}\right)}{\left(\frac{\Omega^2}{g^2}-1\right)\rme^{-\frac{\beta\Delta}{2}}+\cosh\left(\frac{\beta\theta}{2}\right)-\frac{\Delta}{\theta}\sinh\left(\frac{\beta\theta}{2}\right)},
\end{equation}
and
 \begin{equation}\label{eq:theta}
	\theta=\sqrt{\Delta^2+4\Omega_r^2},
\end{equation}
is the polariton detuning in the non-resonant theory. Eq.~\eqref{eq:theta} shows that resonance requires $2\Omega_r\gg\Delta$. Since both $\mathfrak{B}$ and $\Delta$ are functions of $\bar{G}$, one must solve Eqs.~\eqref{eq:OptRes}, \eqref{eq:Bcont}, and \eqref{eq:Delta} self-consistently. 

The value of $\bar{G}$ in Eq.~\eqref{eq:Gbar} and its dependence on the renormalized collective light--matter coupling $\Omega_r$ is essential to understanding how the rates and Lamb shifts scale with the number of molecules. The $N$ scaling of $\bar{G}$ only depends on the size of $\Omega_r$ compared to the temperature,
\begin{equation}
	\Omega_\beta=\frac{10}{\beta}.
\end{equation}
There are slight variations in the $N$ dependence of $\Delta$ and $\mathfrak{B}$ if $\omega_0\lesssim\Omega_\beta$ or $\omega_0\gtrsim\Omega_\beta$, but, these differences do not qualitatively change the $N$ dependence of the master equation. Here in the main text, we present analysis for $\omega_0\lesssim\Omega_\beta$, but, as we show explicitly in Appendix~\ref{app:LowT}, our main conclusions hold when $\omega_0\gtrsim\Omega_\beta$ because the scaling of $\bar{G}$ with $N$ is unchanged. For the typical molecular parameters introduced earlier, one requires temperatures below $7$~K to enter the $\omega_0\gtrsim\Omega_\beta$ regime, and so most molecular experiments are within $\omega_0\lesssim\Omega_\beta$.

\begin{figure}[ht!]\centering
	\includegraphics[width=\columnwidth]{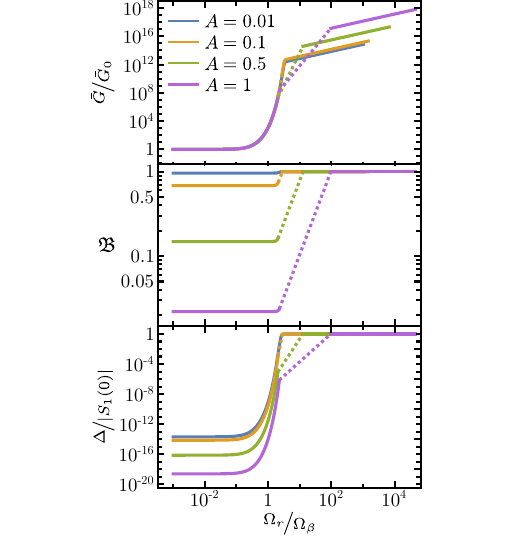}
	\caption{The variational parameter $\bar{G}$, light--matter coupling renormalisation $\mathfrak{B}$, and detuning $\Delta$ are plotted against $\Omega_r/\Omega_\beta$ and vibrational coupling strength $A$ in the experimentally relevant regime with $\omega_0<\Omega_\beta$. The dotted lines are discontinuities in both $x$ and $y$ coordinates, occurring because $\mathfrak{B}$ is discontinuous near $\Omega_r=\Omega_\beta$ and $\mathfrak{B}$ is used in the definition of the $x$-axis. We have used $p=3$ in the spectral density in Eq.~\eqref{eq:J}. Other parameters: the typical molecular values discussed previously. } \label{fig:kappa}
\end{figure}

In Fig.~\ref{fig:kappa} we show $\bar{G}$, $\mathfrak{B}$, and $\Delta$ as a function of $\Omega_r/\Omega_\beta$. There are two distinct regimes, demarcated by $\Omega_r\sim\Omega_\beta$ and with a transitory region near the boundary. One can show that
\begin{equation}\label{eq:GN}
	\bar{G}=\begin{cases}
		\bar{G}_0&\text{ if }\Omega_r\lesssim\Omega_\beta\\
		\Omega_r&\text{ if }\Omega_r\gtrsim\Omega_\beta,
	\end{cases}
\end{equation} 
where \begin{equation}\label{eq:G0}
	\bar{G}_0= \frac{\Delta}{1+\left(\frac{\Delta^2}{g^2\mathfrak{B}^2}-\beta\Delta\right)n_B(\Delta)}.
\end{equation}
Eq.~\eqref{eq:G0} is independent of $N$ which occurs because of the dark state contribution $(\Omega^2/g^2-1)\rme^{-\beta\Delta/2}$ in the denominator of Eq.~\eqref{eq:Gbar}. Without dark states one would instead find that $\lim_{\Omega_r\to 0}\bar{G}\propto N$.

Recent experiments \cite{quach2020organic,shalabney2015coherent} have $\Omega_r<4.2~\text{meV}$ and $20.7~\text{meV}$, and so for room temperature experiments where $\Omega_\beta=258~\text{meV}$, the relevant regime is $\Omega_r\lesssim\Omega_\beta$. For typical molecular parameters, one finds that $\bar{G}_0\approx 2.6\times 10^{-14}~\text{eV}$. Since the integrand of $\Delta$ in Eq.~\eqref{eq:Delta} scales with $(1-G(\omega))^2\propto\bar{G}^2$, $\Delta$ is negligible in this regime---as also shown in Fig.~\ref{fig:kappa}---and so for typical molecular parameters one can safely take the resonant limit of the variational polaron theory. The resonant value of $\bar{G}_0$ is
\begin{equation}\label{eq:Gres}
	\bar{G}_0=\frac{2 g_r^2\beta}{2+g_r^2\beta^2}.
\end{equation}
The small value of $\bar{G}_0$ for typical molecular parameters means that the WCME is expected to be very inaccurate for $\Omega_r\lesssim\Omega_\beta$. Moreover, in this regime, Fig.~\ref{fig:kappa} shows that the light--matter coupling can be heavily suppressed by $\mathfrak{B}$ for strong vibrational coupling. 

In the regime less relevant to recent experiments \cite{quach2020organic,shalabney2015coherent}, $\Omega_r\gtrsim\Omega_\beta$, we find $\bar{G}\propto\Omega_r$ with the proportionality becoming an equality if the system is approximately resonant, $2\Omega_r\gg\Delta$. In this very strong light--matter coupling regime (or very low temperature regime), $\bar{G}$ may be large enough that $G(\omega)\approx 0$ for all phonon frequencies $\omega$ that contribute to the dynamics. In this regime the WCME will be accurate and $\mathfrak{B}\approx 1$. However, also in this regime, Fig.~\ref{fig:kappa} shows that the detuning becomes equal to the vibrational reorganization energy. Consequently, one must assess whether the system is resonant by comparing the size of $2\Omega_r$ to $\Delta=-S_1(0)$. 

In Table~\ref{tab:regimes} we summarize the $\omega_0\lesssim\Omega_\beta$ regime of the variational transformation.  Until Section~\ref{sec:NonResonance}, we now enforce resonance, $\Delta\equiv 0$, because this describes the most experimentally relevant parameter regimes, and substantially simplifies the presentation of equations. In Section~\ref{sec:NonResonance} we summarize important corrections in the non-resonant regime.

 \begin{center}
 	\begin{table}[]
 		\begin{tabular}{@{}ccc@{}}
 			\toprule
 			\multicolumn{3}{c}{$\omega_0 \lesssim \Omega_\beta$} \\ \midrule
 			\multirow{2}{*}{$\Omega_r\lesssim\Omega_\beta$} & \multicolumn{2}{c}{$\Omega_r\gtrsim\Omega_\beta$} \\ \cmidrule(l){2-3} 
 			& $2\Omega_r\lesssim|S_1(0)|$ & $2\Omega_r\gg |S_1(0)|$ \\ \midrule
 			$\bar{G}=\bar{G}_0$ & $\bar{G}\propto \Omega_r$ & $\bar{G}= \Omega_r$ \\
 			$G(\omega)\approx 1$ & $G(\omega)\approx 0$ & $G(\omega)\approx 0$ \\
 			$\Delta=0$ & $\Delta\neq0$ & $\Delta= 0$ \\ \bottomrule
 		\end{tabular}
 		\caption{The variational polaron transformation for the different parameter regimes. The experimentally relevant regime has both $\omega_0\lesssim\Omega_\beta$ and $\Omega_r\lesssim\Omega_\beta$. $G(\omega)$ and $\Delta$ values shown are for the typical molecular parameters. For a similar breakdown in the regime $\omega_0\gtrsim\Omega_\beta$, see Appendix~\ref{app:LowT}.}
 		\label{tab:regimes}
 	\end{table}
 \end{center}

\section{The variational polaron master equation}\label{sec:VPME}

We are now in a position to derive the Redfield master equation in the variational polaron frame. As a second order perturbation in $\mc{H}_{SB}$ given in Eqs.~\eqref{eq:VE}--\eqref{eq:VP} we expect the variational polaron master equation (VPME) to have three distinct contributions. First, a displacement--type master equation $\mc{O}(\mc{H}_D^2)$, which by comparison of Eq.~\eqref{eq:VE} to Eq.~\eqref{eq:HSB} will be identical to the WCME in Eq.~\eqref{eq:MEweak} but with the replacement $J(\omega)\to J(\omega)(1-G(\omega))^2$ in the Fourier transforms of the correlation functions. Second, a polaron--type master equation $\mc{O}(\mc{H}_{P}^2)$, and finally, a variational--type contribution $\mc{O}(\mc{H}_D\mc{H}_{P})$.

Rather than obfuscating the text by stating the general non-secular VPME---which we give in Appendix~\ref{app:ME}---we will instead move onto discussing the transition rates, dephasing rates, and Lamb shifts predicted by the VPME. Analagously to the WCME in Eq.~\eqref{eq:dotrho}, the relevant part of the VPME is
\begin{equation}\label{eq:VPME}
	\dot{\varrho}_{\mu\nu}(t)= -R_{\mu\nu} \varrho_{\mu\nu}(t)+\ldots,
\end{equation}
where $\varrho_{\mu\nu}(t)=\braket{\mu|\varrho(t)|\nu}$ is an element of the variational frame density operator $\varrho(t)=U\rho(t)U^\dagger$ and
\begin{equation}\label{eq:Rmunu}
	R_{\mu\nu}=\frac{K_\mu^\downarrow+K_{\nu}^\downarrow}{2}+K^\phi_{\mu\nu}+i\Delta_{\mu\nu},
\end{equation}
where capitalized symbols are the equivalent quantities in the VPME to the corresponding lowercase symbols in Eq.~\eqref{eq:Gmunu} for the WCME. The loss rates can be written as summations of the transition rates,
\begin{equation}
		K_\mu^\downarrow=\sum_{\alpha\neq\mu}K_{\mu\to\alpha},
\end{equation}
and the Lamb shifted transition frequencies are
\begin{equation}
\Delta_{\mu\nu}=\left(\omega_\mu+\Lambda_\mu\right)-\left(\omega_\nu+\Lambda_\nu\right).
\end{equation}

\subsection{Transition rates}\label{sec:transition}
The displacement--type and variational--type master equations generate single phonon processes, whilst the polaron--type master equation generates single and multi phonon processes. After collecting all terms, one finds that the transition rates are
\begin{align}
	K_{\pm\to\mp}&=\frac{1}{4N}\gamma(\pm 2\Omega_r),\label{eq:gv1}\\
	K_{\pm\to d}&=K_{d\to\mp}=\frac{1}{2N}\gamma(\pm\Omega_r),\label{eq:gv2}
\end{align} 
and $K_{d\to d \neq d'}=k_{d\to d\neq d'}$ is equal to the WCME rate in Eq.~\eqref{eq:gw4}. We have defined $\gamma(\nu)=2\text{Re}[\Gamma(\nu)]$ with 
\begin{equation}
	\Gamma(\nu)=\Gamma_1(\nu)+\Gamma_{>1}(\nu),
\end{equation} 
where $\Gamma_1(\nu)$ is the Fourier transform of the single phonon bath correlation function, with real and imaginary parts given in Eqs.~\eqref{eq:g1} and \eqref{eq:S}, respectively. The Fourier transform of the multi phonon bath correlation function is
\begin{equation}\label{eq:Gg1}
\Gamma_{>1}(\nu)=\nu^2\sum_{\substack{m\in \{m(\nu)\}}} \frac{1}{m!}\Phi_m(\nu),
\end{equation}
with $\{m(\pm\Omega_r)\}=\{2,3,4,\ldots\}$ and $\{m(\pm2\Omega_r)\}=\{3,5,7,\ldots\}$, and we have defined
\begin{equation}\label{eq:Phin}
\Phi_m(\nu)=\inta{\tau}\rme^{i\nu\tau}\phi(\tau)^m,
\end{equation}
as the Fourier transform of the $m$th power of the phonon propagator,
\begin{multline}\label{eq:phi}
\phi(\tau)=\\\inta{\omega}J(\omega)\frac{G(\omega)^2}{\omega^2}\left(n_B(\omega)\rme^{i\omega s}+\tilde{n}_B(\omega)\rme^{-i\omega s}\right).
\end{multline}
The $m$th term in the summation in Eq.~\eqref{eq:Gg1} is the contribution of processes involving $m$ phonons.  The total loss rates from the eigenstates are
\begin{align}
	K_\pm^\downarrow&=\frac{1}{4N}\gamma(\pm 2\Omega_r)+\frac{N-1}{2N}\gamma(\pm\Omega_r),\\
	K_d^\downarrow&=\frac{1}{2N}\left[\gamma(\Omega_r)+\gamma(-\Omega_r)\right]+\frac{N-2}{N}\gamma_1(0).
\end{align}

Compared to transition rates in the WCME in Eqs.~\eqref{eq:kpmw}--\eqref{eq:kdw}, transition rates predicted by the VPME differ in two important ways. First, by the replacement $\Omega\to\Omega_r$ in the transition energies, the effects of which are well demonstrated in Fig.~\ref{fig:kappa}. Second, by the introduction of multi phonon transitions. 

\begin{figure}[ht!]\centering
	\includegraphics[width=\columnwidth]{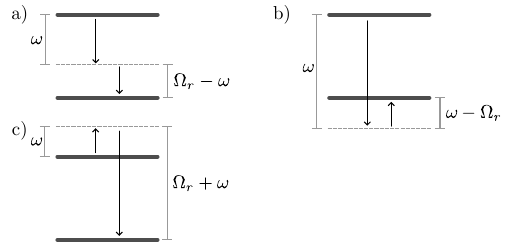}
	\caption{The transitions described by the integrands in Eq.~\eqref{eq:2phonon}. The downwards and upwards arrows denote emission and absorption of a phonon of the indicated energy, respectively. (a), (b) and, (c) correspond to the first, second, and third terms of Eq.~\eqref{eq:2phonon}, respectively, and $\omega$ is integrated between the limits shown in Eq.~\eqref{eq:2phonon}. Since the transition energy is $\Omega_r$, the upper and lower states could be $\ket{+}$ and $\ket{d}$, or $\ket{d}$ and $\ket{-}$.} \label{fig:2phonon}
\end{figure}

The effect of multi phonon transitions depends on the size of $\Omega_r$ compared to the high frequency cutoff $\omega_0$. To demonstrate why, it is helpful to expand the bath correlation function associated with two phonon decay into its contributions. For decay processes at transition frequency $\Omega_r$, the two-phonon contributions to the rates are proportional to
\begin{multline}\label{eq:2phonon}
	\frac{1}{\pi}\text{Re}[\Phi_2(\Omega_r)]=\\\int_0^{\Omega_r}\ \text{d}\omega\  J_P(\omega)J_P(\Omega_r-\omega)\tilde{n}_b(\omega)\tilde{n}_b(\Omega_r-\omega)\\
	+\int_{\Omega_r}^\infty \text{d}\omega\  J_P(\omega) J_P(\omega-\Omega_r)\tilde{n}_b(\omega)n_b(\omega-\Omega_r)\\
	+\int_0^\infty  \text{d}\omega\  J_P(\omega)J_P(\Omega_r+\omega)n_b(\omega)\tilde{n}_b(\Omega_r+\omega),
\end{multline}
where $J_P(\omega)=J(\omega)G(\omega)^2/\omega^2$. In Fig.~\ref{fig:2phonon} we illustrate the transitions described by Eq.~\eqref{eq:2phonon}. 

 When $\Omega_r \gg \omega_0$, the transition rates are significantly suppressed by the high frequency cutoff of the vibrational baths. In this regime, multi phonon processes dominate over single phonon processes, because the process of decaying by emitting two phonons of energy less than $\Omega_r$---shown in Fig.~\ref{fig:2phonon}(a)---is substantially more probable than emitting a single phonon of energy $\Omega_r$. This is because $J(\Omega_r)\ll J(\Omega_r/2)^2$ when $\Omega_r\gg\omega_0$ for typical spectral densities \footnote{Multi phonon transitions will only dominate in the regime $\Omega_r\gg\omega_0$ for spectral densities satisfying $J(\omega)\ll J(\omega/n)^n$. For example, if $J(\omega)\propto \omega^p\exp(-(\omega/\omega_0)^q)$, this holds for $q>1$. We argue that most real spectral densities satisfy this property.}. The same arguments apply for decays and excitations through higher order phonon processes; however, unless the vibrations are very strong ($A \gg 1$), processes involving more than two or three phonons will not significantly contribute.
 
 On the other hand, when $\Omega_r \lesssim \omega_0$, the cut-off frequency of the bath does not have as great an effect on the rates. Whether or not multi phonon processes are dominant in this regime, and how they scale with $N$, depends on the form of the spectral density $J(\omega)$. In general, one must evaluate the rate functions in Eqs.~\eqref{eq:gv1} and \eqref{eq:gv2} to determine the contribution of multi phonon processes when $\Omega_r\lesssim\omega_0$. 
 
\begin{figure*}[ht!]\centering
 	\includegraphics[width=\textwidth]{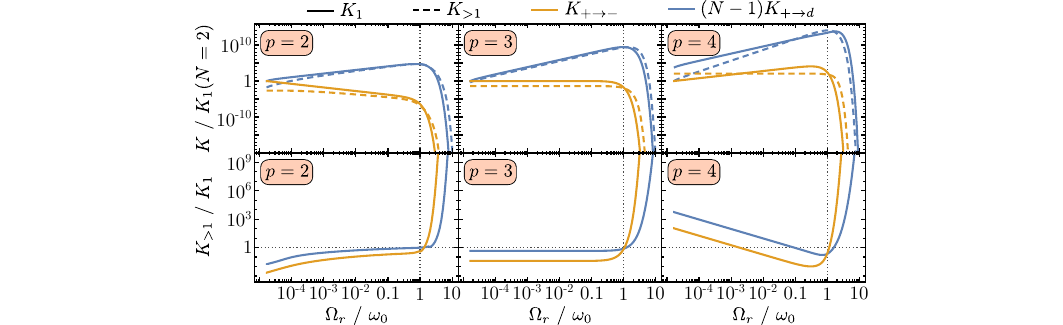}
 	\caption{The transition rates from $\ket{+}$ to $\ket{-}$ ($K_{+\to -})$ and from $\ket{+}$ to all dark states $\ket{d}$ ($(N-1)K_{+\to d}$) as a function of $\Omega_r$. In the top row we show the single phonon ($K_1$---solid curves) and multi phonon ($K_{>1}$---dashed curves) contributions keeping up to third order phonon interactions and normalized with respect to the single phonon contribution when $N=2$. In the bottom row we plot the ratio of the single and multi phonon contributions shown in the panel above. Each column has a different Ohmicity, $p$, in the spectral density in Eq.~\eqref{eq:J}. This figure shows that when $\Omega_r\lesssim\omega_0$, whether multi phonon contributions dominate depends on the particular model, and that when $\Omega_r\gtrsim\omega_0$ multi phonon contributions always dominate. Parameters common to all panels: $\omega_0=6~\text{meV}$, $g=0.1~\mu\text{eV}$, and $T=300~\text{K}$. For Ohmicity $p=3$ and $p=4$ we use $A=0.083$ whilst for $p=2$ we use $A=0.0083$, which give a similar value of $\mathfrak{B}$ in each column. } \label{fig:TransitionRates}
 \end{figure*}

 In Fig.~\ref{fig:TransitionRates} we calculate the multi phonon contributions for Ohmicities $p=2$, $p=3$, and $p=4$ ($p$ is defined in Eq.~\eqref{eq:J}) in the experimentally relevant regime $\Omega_r\lesssim\Omega_\beta$. One can see that each Ohmicity affords different ratios of single to multi phonon contributions when $\Omega_r\lesssim\omega_0$, but in all cases the multi phonon contributions strongly dominate when $\Omega_r\gg\omega_0$. 

 This discussion holds for all values of $\Omega_r$ compared to $\Omega_\beta$ so long as the system is resonant to a good approximation. In the regime less relevant to recent experiments, $\Omega_r\gtrsim\Omega_\beta$, the $m$ phonon contribution scales by a factor of $N^{-m/2}$ differently to the same contribution when $\Omega_r\lesssim\Omega_\beta$ because, when $\Omega_r\gtrsim\Omega_\beta$, $\bar{G}\propto\Omega_r$ as shown in Eq.~\eqref{eq:GN}. In the regime $\Omega_r\gtrsim\omega_0$, the additional factor of $N^{-m/2}$ will not change the fact that multi phonon processes will be exponentially more probable than single phonon processes. 
 
 This leads us to our first main result illustrated in Fig.~\ref{fig:sketch}(b). In the large $N$ limit the dark states become population sinks, and, if $\Omega_r\gtrsim\omega_0$, the transitions are strongly dominated by multi phonon processes.

\subsection{Dephasing rates}\label{sec:decoherence}
The dephasing rates have contributions from the displacement--type and polaron--type master equations, but not from the variational--type master equation. One finds that the overall dephasing rate of the coherence between states $\ket{\mu}$ and $\ket{\nu}$ is
\begin{equation}
K_{\mu\nu}^\phi=k^{\phi}_{\mu\nu}+k^{\Phi}_{\mu\nu},	
\end{equation}
where $k^{\phi}_{\mu\nu}$ is the single phonon contribution arising from the displacement--type interaction, exactly equal to the WCME dephasing rate in Eq.~\eqref{eq:gWPhi}, and 
\begin{equation}\label{eq:gPphi}
	k^{\Phi}_{\mu\nu}=\gamma^\phi_{>1}(0)\left[\frac{\delta_{\mu\pm}+\delta_{\nu\pm}}{2}-\mu\nu\delta_{\mu\pm}\delta_{\nu\pm}\right],
\end{equation}
is the polaron--type dephasing. The first term of $k_{\mu\nu}^{\Phi}$ is the contribution from virtual self transitions whilst the second term is from other pure dephasing processes. The function $\gamma_{>1}^\phi(0)=2\text{Re}[\Gamma_{>1}^\phi(0)]$ where
\begin{equation}\label{eq:gP0}
	\Gamma^\phi_{>1}(0)=g^2\mathfrak{B}^2\inta{\tau}\left(\cosh\left(\phi(\tau)\right)-1\right),
\end{equation} 
describes multi phonon dephasing processes. Using Eqs.~\eqref{eq:gWPhi} and \eqref{eq:gPphi} we find the following dephasing rates of the coherences,
\begin{align}
	K_{+-}^\phi&=2\gamma^\phi_{>1}(0),\label{eq:gp1}\\
K^
	\phi_{\pm G}&=K^\phi_{\pm d}=\frac{1}{8N}\gamma_1(0)+\frac{1}{2}\gamma^\phi_{>1}(0),\label{eq:gp2}\\
	K_{dG}^\phi&=\frac{1}{2N}\gamma_1(0)\label{eq:gp3},\\
	K_{d_id_j}^\phi&=0.\label{eq:gp4}
\end{align}
Notice that multi phonon dephasing contributes to $K_{+-}^\phi$---whereas for symmetry reasons single phonon dephasing did not---and to $K_{\pm G}^\phi$ which contributes to the polariton line widths.

By expanding $\cosh(\phi(\tau))$ in Eq.~\eqref{eq:gP0} as a series in $\phi(\tau)$, one finds that polaron--type dephasing is caused by processes with even numbers of phonons. The lowest order contribution is second order,
\begin{multline}
	\gamma^\phi_{>1}(0)=2\pi g^2\mathfrak{B}^2\inta{\omega}\left[\frac{J(\omega)G(\omega)^2}{\omega^2}\right]^2\\ \times n_B(\omega)\left[1+n_B(\omega)\right]+\ms{O}(J(\omega)^4),
\end{multline}
which describes simultaneous phonon absorption and emission at all possible frequencies. The leading order contribution of the multi phonon dephasing has the following scaling with $N$,
\begin{equation}\label{eq:gg1scale}
	\gamma_{>1}^\phi(0)\sim\begin{cases}
		1&\text{ if }\Omega_r\lesssim\Omega_\beta,\\
		\frac{1}{N^2}&\text{ if }\Omega_r\gtrsim\Omega_\beta,
	\end{cases}
	\end{equation} 
which follows from Eq.~\eqref{eq:GN}.

In the variational polaron frame, the absorption spectrum of the cavity is described by Eq.~\eqref{eq:A2} but using the quantities from the VPME. For instance, the line widths of the polariton peaks are equal to $2\text{Re}[R_{\pm G}]=K_\pm^\downarrow+2 K_{\pm G}^\phi$. In the experimentally relevant regime, $\Omega_r\lesssim\Omega_\beta$, one finds from Eqs.~\eqref{eq:gp2} and \eqref{eq:gg1scale} that the leading order contribution to the line widths is multi phonon dephasing independent of $N$. Conversely, when $\Omega_r\gtrsim\Omega_\beta$, the leading order contribution is single phonon dephasing scaling as $\gamma_1(0)/N$; however, as discussed in Section~\ref{sec:weak}, $\gamma_1(0)$ is zero or divergent for many spectral density choices. The next leading order term is two-phonon dephasing scaling as $1/N^2$. 

Although less relevant to recent experiments \cite{quach2020organic,shalabney2015coherent}, it is important to understand the leading order dephasing rate when $\Omega_r\gg\Omega_\beta$. To do so we must clarify the zeros and divergences in the single phonon contribution, $\gamma_1(0)$, given in Eq.~\eqref{eq:g10}. As discussed in Section~\ref{sec:weak}, these non-finite results stem from an unjustified Markovian assumption. Specifically, taking the infinite limit of the upper integration domain in the bath correlation function. Relaxing this assumption one finds that the single phonon pure dephasing rate is 
\begin{multline}\label{eq:g10t}
	\gamma_1(0,\tau)=2\inta{\omega}J(\omega)(1-G(\omega))^2\\\times\coth\left(\frac{\beta\omega}{2}\right)\frac{\sin\left(\omega \tau\right)}{\omega}.
\end{multline}
The Markovian limit is recovered by using the $\delta$-function representation: $\lim_{\tau\to\infty}\sin(\omega \tau)/\omega=\pi\delta(\omega)$. 

Eq.~\eqref{eq:g10t} should replace the factors of $\gamma_1(0)$ appearing in Eqs.~\eqref{eq:gp2}--\eqref{eq:gp3}. After a time $t$, the non-Markovian single phonon rate $\gamma_1(0,\tau)$ suppresses the coherences by a factor of
\begin{align}\label{eq:decF}
	D_{a,N}(t)&=\exp\left(-\frac{1}{a N}\int_0^t\text{d}\tau\ \gamma_1(0,\tau)\right)\nonumber\\
 &=\exp\left(-\frac{1}{a N}\text{Re}\left[\phi_D(t)-\phi_D(0)\right]\right),
\end{align} 
compared to the initial value of the coherences. The factor of $1/(aN)$ in the exponent of Eq.~\eqref{eq:decF}, where $a$ is constant, is the coefficient of $\gamma_1(0,\tau)$ in either Eq.~\eqref{eq:gp2} or Eq.~\eqref{eq:gp3}. For example, $D_{8,N}(t)$ is the suppression factor of $\varrho_{\pm G}(t)$ and $\varrho_{\pm d}(t)$ at time $t$ due to $\gamma_1(0,\tau)$ appearing in Eq.~\eqref{eq:gp2}. In Eq.~\eqref{eq:decF} we have also defined the displacement--type phonon propagator,
\begin{multline}\label{eq:phiD}
\phi_D(t)=\inta{\omega}\frac{J(\omega)\left(1-G(\omega)\right)^2}{\omega^2}\\\times\left(n_B(\omega)\rme^{i\omega s}+\tilde{n}_B(\omega)\rme^{-i\omega s}\right),
\end{multline} 
which is analogous to the polaron--type phonon propagator in Eq.~\eqref{eq:phi}. The exponent of Eq.~\eqref{eq:decF} is the so-called decoherence function \cite{breuer2002theory} which describes pure dephasing caused by emitting a finite frequency phonon at time $0$ and re-absorbing the same phonon at time $t\ge 0$. Note that $D_{1,1}(t)$ \textit{exactly} describes the dephasing of the excited--ground state coherence of an isolated molecule with Hamiltonian $H_\text{mol}$ in Eq.~\eqref{eq:Hmol} \cite{breuer2002theory}.

\begin{figure}[ht!]\centering
	\includegraphics[width=\columnwidth]{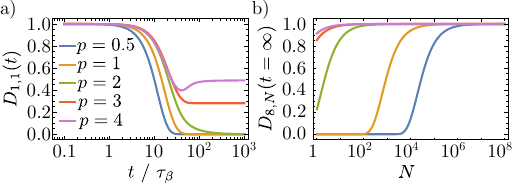}
	\caption{(a) $D_{1,1}(t)$ is shown as a function of time, and (b) $D_{8,N}(t=\infty)$ as a function of $N$. In both panels, the different coloured curves correspond to different Ohmicity values $p$ of the spectral density in Eq.~\eqref{eq:J}, indicated in the legend, and  $\tau_\beta=\beta/\pi$ is the thermal bath time \cite{breuer2002theory}. We enforce $G(\omega)=0$ for all parameters shown because we are interested only in the regime with $\Omega_r\gtrsim\Omega_\beta$. Other parameters take the typical molecular values introduced earlier.} \label{fig:DaN}
\end{figure}

As shown in Fig.~\ref{fig:DaN}(a), $D_{1,1}(t)$ describes approximate exponential suppression of the coherences in time, but, crucially, to a non-zero---albeit sometimes small---minimum value. This is an important distinction between Markovian and non-Markovian dephasing. On the one hand, Markovian dephasing---with a generic time-independent rate  $\gamma_M$---leads to a coherence suppression factor at time $t$ of $\exp(-\gamma_M t)$. On the other hand, non-Markovian dephasing---with a generic time-dependent rate $\gamma_{nM}(\tau)$---leads to a suppression factor of $\exp[-\int_0^t\text{d}\tau\ \gamma_{nM}(\tau)]$. Therefore, whilst Markovian dephasing always suppresses the coherences to zero when $t\to\infty$, the non-Markovian suppression factor may be greater than zero at $t\to\infty$. This difference is particularly important for molecular experiments, where the regime in which $\gamma_1(0,\tau)$ is relevant, $\Omega_r\gtrsim\Omega_\beta$, can only be reached with $N\gtrsim 10^{12}$ molecules. Fig.~\ref{fig:DaN}(b) shows that the $1/N$ suppression within the exponent of $D_{a,N}(t)$ causes the long-time limit to approach unity, even for $N\ll 10^{12}$. Indeed, one finds that $D_{a,N}(t)\to 1$ at all times for relatively small values of $N$, such that single phonon dephasing becomes negligible. This discussion suggests that all non-Markovian contributions to rates that scale inversely with $N$---which includes all rates in the VPME with the exception of $\gamma_{>1}^\phi(0)$ in the regime $\Omega_r\lesssim\Omega_\beta$---are negligible compared to the Markovian contributions.

This finding leads us to our second main result, illustrated in Fig.~\ref{fig:sketch}(c). When $\Omega_r\gtrsim\omega_0$, decoherence is dominated by dephasing, which, for typical molecular parameters, is a multi phonon processes involving all even orders of phonon interactions that scales independently of $N$ when $\Omega_r\lesssim\Omega_\beta$, and a two phonon process scaling as $1/N^2$ when $\Omega_r\gtrsim\Omega_\beta$.

\subsection{Lamb shifts}\label{sec:Lamb}
We find the following expressions for the Lamb shifts,
\begin{align}
	\Lambda_\pm&=\frac{1}{4N}S^v(\pm 2\Omega_r)+\frac{N-1}{2N}S^v(\pm\Omega_r)\nonumber\\
 &\quad+\frac{1}{4N}S^v_1(0)+S_{>1}^\phi(0),\label{eq:Lv1}\\
	\Lambda_d&=\frac{1}{2N}\left[S^v(\Omega_r)+S^v(-\Omega_r)\right]+\frac{N-2}{N}S_1^v(0)\nonumber\\
 &\quad+\frac{1}{N}S^v_1(0),\label{eq:Lv2}
\end{align}
and $\Lambda_G=0$, where we have defined $S^v(\nu)=S_1^v(\nu)+S_{>1}(\nu)$ which has the single phonon contribution
\begin{multline}\label{eq:Sv}
	S_1^v(\nu)=\mathcal{P}\inta{\omega}J(\omega)\Big[\frac{\tilde{n}_B(\omega)}{\nu-\omega}\mathcal{G}(\nu,\omega)\\+\frac{n_B(\omega)}{\nu+\omega}\mathcal{G}(-\nu,\omega)\Big],
\end{multline}
with
\begin{equation}
\mathcal{G}(\nu,\omega)=\left(1+\left[\frac{\nu}{\omega}-1\right]G(\omega)\right)^2,
\end{equation}
and multi phonon contribution $S_{>1}(\nu)=\text{Im}[\Gamma_{>1}(\nu)]$ where $\Gamma_{>1}(\nu)$ is given in Eq.~\eqref{eq:Gg1}. Eq.~\eqref{eq:Lv1} also contains a contribution from virtual multi phonon transitions, $S_{>1}^\phi(0)=\text{Im}[\Gamma_{>1}^\phi(0)]$, where $\Gamma_{>1}^\phi(0)$ is given in Eq.~\eqref{eq:gP0}. 

The first lines of Eqs.~\eqref{eq:Lv1}--\eqref{eq:Lv2} arise from transitions between states, with the $(N-2)S_1^v(0)/N$ contribution in Eq.~\eqref{eq:Lv2} arising from transitions between degenerate dark states. The second lines of Eqs.~\eqref{eq:Lv1}--\eqref{eq:Lv2} arise from virtual transitions from a state back to itself.
 
In the large $N$ limit, the Lamb shifts in Eqs.~\eqref{eq:Lv1}--\eqref{eq:Lv2} are dominated by contributions from single phonon transitions involving the large numbers of dark state, such that
\begin{align}
    \Lambda_\pm&\approx\frac{N-1}{2N}S_1^v(\pm\Omega_r),\\
    \Lambda_d&\approx\frac{N-2}{N}S_1^v(0)\label{eq:S1v0}.
\end{align}

In the WCME discussed in Section~\ref{sec:LambWeak}, we found that the polariton shifts were negligible compared to the Rabi splitting because, in the WCME, when $|\nu|\gg\omega_0$ the relevant function $S_1(\nu)$ in Eq.~\eqref{eq:S} was inversely dependent on $\nu$. Consequently, $S_1(\pm\Omega)$ and $S_1(\pm2\Omega)$ vanished in the large $N$ limit. However, in the variational polaron theory, the relevant function is instead $S_1^v(\nu)$ in Eq.~\eqref{eq:S1v}, which generally has a different dependence on frequency owing to the presence of $\ms{G}(\pm\nu,\omega)$. We are able to evaluate the $S_1^v(\nu)$ when $|\nu|\gg\omega_0$ in the limits $\Omega_r\lesssim\Omega_\beta$ and $\Omega_r\gtrsim\Omega_\beta$, because, for typical molecular parameters, Eq.~\eqref{eq:GN} indicates that we can approximate $G(\omega)=1$ and $G(\omega)=0$ in these limits, respectively. Within these parameter regimes one finds that
\begin{equation}\label{eq:Lpm}
    \lim_{\Omega_r\gg\omega_0}\Lambda_\pm\approx\begin{cases}
        \pm\frac{\Omega_r}{2}B_2(\beta)&\text{ if }\Omega_r\lesssim\Omega_\beta\\
        \pm\frac{1}{2\Omega_r}B_0(\beta) &\text{ if }\Omega_r\gtrsim\Omega_\beta,
    \end{cases}
\end{equation}
where
\begin{equation}
    B_j(\beta)=\inta{\omega}\frac{J(\omega)}{\omega^j}\coth\left(\frac{\beta\omega}{2}\right).
\end{equation}
Eq.~\eqref{eq:Lpm} shows that in the parameter regime relevant to recent experiments \cite{quach2020organic,shalabney2015coherent}, $\Omega_r\lesssim\Omega_\beta$, single phonon Lamb shifts modify the polariton energies to 
\begin{equation}    \label{eq:LambPM}\omega_\pm\to\omega_c\pm\Omega_r\left(1+\frac{1}{2}B_2(\beta)\right).
\end{equation}
For example, for the typical molecular parameters at room temperature, $B_2(\beta)=0.634$ when $p=3$, which is a significant Lamb shift \footnote{For $p<2$ $B_2(\beta)$ diverges but this is because we have approximated $\bar{G}=0$ to arrive at Eq.~\eqref{eq:Lpm}, whilst in fact it is non-zero but small.}. 

Returning to the leading order contribution to the dark state Lamb shift in Eq.~\eqref{eq:S1v0}, using Eq.~\eqref{eq:Sv} one finds that in the large $N$ limit
 \begin{equation}\label{eq:S1v}
 	\Lambda_d\approx -\inta{\omega}\frac{J(\omega)}{\omega}\left(1-G(\omega)\right)^2=-\Delta,
 \end{equation}
 which is equal to the negative of the detuning in $\ms{H}_S$. This Lamb shift is negligible for typical molecular parameters in the limit $\Omega_r\lesssim\Omega_\beta$ because $1-G(\omega)\propto \bar{G}$, but may be large when $\Omega_r\gtrsim\Omega_\beta$ if the system is non-resonant. 

 This discussion brings us to our third main result, illustrated in Fig.~\ref{fig:sketch}(d). When $\Omega_r\gtrsim\omega_0$ and for typical molecular parameters, if $\Omega_r\lesssim\Omega_\beta$ the polaritons are Lamb shifted by equal and opposite amounts proportional to $\Omega_r$, whilst, if $\Omega_r\gtrsim\Omega_\beta$, the dark state is Lamb shifted by an amount equal to the vibrational reorganization energy.

 \section{Non-resonance}\label{sec:NonResonance}
We now briefly summarize the effects of non-resonance on our three main conclusions summarized in Fig.~\ref{fig:sketch} regarding multi phonon transitions, dephasing, and Lamb shifts. We derive these results from the non-resonant variational polaron master equation given in Appendix~\ref{app:resonance}. Recall that inadvertent non-resonance occurs if both $|S_1(0)|\gtrsim 2\Omega_r$ and $\Omega_r\gtrsim\Omega_\beta$ are satisfied, which requires strong vibrational coupling and either strong light--matter coupling or very low temperature. One could also avoid inadvertent non-resonance by building a wedge-shaped cavity and continuously tuning the mode energy until it becomes resonant with the matter system \cite{schutte2008continuously}.

Regarding multi phonon transitions into the dark states, when the system is non-resonant, transitions between the polariton states and the dark states become
\begin{equation}\label{eq:KD}
K_{\pm\to d}(\Delta)=\left(1\pm\frac{\Delta}{\theta}\right)K_{\pm\to d}(0),
\end{equation}
where $\theta$ is the polariton detuning in Eq.~\eqref{eq:theta} and $K_{\pm\to d}(0)$ are the resonant values in Eq.~\eqref{eq:gv2}. Consequently, in far off-resonant systems where $\Delta$ is comparable or larger than $2\Omega_r$, transitions from the upper and lower polaritons to the dark states will be, respectively, enhanced and suppressed.

Regarding multi phonon dephasing, the non resonant expression is
\begin{equation}\label{eq:gD}
	\gamma_{>1}^\phi(\Delta,0)=\left(1-\frac{\Delta^2}{\theta^2}\right)\gamma_{>1}^\phi(0),
\end{equation} 
where $\gamma_{>1}^\phi(0)$ is the resonant multi phonon dephasing rate defined through Eq.~\eqref{eq:gP0}. In far off-resonant systems, the dephasing rate will be slower than anticipated from a resonant theory.

Lastly, non-resonance does not affect Lamb shifts. This is because non-resonance requires $\Omega_r\gtrsim\Omega_\beta$, and in this limit we have shown in Section~\ref{sec:Lamb} that the only non-negligible Lamb shift is to the dark state energy, originating from transitions between degenerate dark states. Since non-resonance does not affect properties of the dark states, this Lamb shift is also unaffected. 

The non-resonant corrections in Eqs.~\eqref{eq:KD}--\eqref{eq:gD} scale to leading order as $[\Delta/(2\Omega_r)]^2$. Since non-resonance requires $\Omega_r\gtrsim\Omega_\beta$, and at room temperature $\Omega_\beta =0.258~\text{eV}$, non-resonant corrections will only appear for very strong vibrational coupling strengths, or in low temperature experiments. For example, at room temperature and for an Ohmic spectral density with $\omega_0=6~\text{meV}$, one requires $A>30$ for $[\Delta/(2\Omega_\beta)]^2>0.1$, which is orders of magnitude larger than the typical molecular value of $A=0.083$. Conversely, for $A=0.083$, one requires $T<0.8~\text{K}$.

\section{Conclusion}\label{sec:conclusion}

By deriving the Redfield equation in the variational polaron frame we have shown that multi phonon processes and vibrational suppression of the light--matter coupling are important phenomena in molecular polaritonics. 

Vibrational displacement interactions cause transitions between the upper polariton, lower polariton, and the dark states. When the collective light--matter coupling is smaller than the high frequency cutoff of the vibrations ($\Omega_r\lesssim \omega_0$) we have shown in Section~\ref{sec:transition} that whether multi phonon processes dominate transition rates depends strongly on the spectral density. Conversely, when $\Omega_r\gg \omega_0$---a parameter regime now accessible to experiments \cite{shalabney2015coherent}---one finds that multi phonon processes always dominate. An important result found in Refs.~\cite{de2018cavity,martinez2018comment}, valid for weak vibrational coupling, was that the dark states act as population sinks when there are a large number of molecules. This result holds at strong vibrational coupling, but we found here that the transfer is carried out through single \textit{and} multi phonon processes when $\Omega_r\lesssim\omega_0$, and almost exclusively by multi phonon processes when $\Omega_r\gg\omega_0$.

Vibrational displacement interactions also cause dephasing of eigenstate coherences. This is particularly important in the limit $\Omega_r\gg\omega_0$ where the contribution of the decay rates to decoherence is exponentially suppressed with increasing $N$ such that decoherence is dominated by dephasing. In Section~\ref{sec:decoherence}, we found that dephasing is always a multi phonon process to leading order in $N$. When $\Omega_r\lesssim\Omega_\beta$---the regime relevant to recent experiments \cite{quach2020organic,shalabney2015coherent}---one finds that dephasing is independent of $N$, whilst when $\Omega_r\gtrsim\Omega_\beta$, dephasing scales as $1/N^2$. The $N$-independence of the dephasing rates when $\Omega_r\lesssim\Omega_\beta$ originates from the contribution of the $N-1$ dark states to the free energy of the system, manifesting in $\bar{G}$ being independent of $N$ in Eq.~\eqref{eq:GN}. This is an important and novel role of dark states in molecular polaritonics.

In Section~\ref{sec:Lamb} we showed that, when $\Omega_r\lesssim\Omega_\beta$, the polariton energies can be significantly Lamb shifted even for only moderately strong vibrational coupling. This prediction cannot be obtained from the weak vibrational coupling theory, and arises due to transitions from the polaritons into the dark states. We derived a simple expression for the Lamb shifted polariton energies, given in Eq.~\eqref{eq:LambPM}, valid for when $\Omega_r\gg\omega_0$. We also showed that if $N$ is increased or the temperature is reduced such that $\Omega_r\gtrsim\Omega_\beta$, then the polariton Lamb shifts become negligible but the dark states become Lamb shifted by an amount equal to the negative of the cavity--molecule detuning.

Finally, in Section~\ref{sec:NonResonance} we briefly discussed corrections to the multi phonon transition and dephasing rates when the model is non-resonant. We found that non-resonant effects are likely to be negligible for molecular experiments unless they are performed at temperatures below one Kelvin.

\bibliographystyle{PRX}
\bibliography{Pinopaperbib}

\clearpage
\appendix

\section{Data from Ref.~\cite{quach2020organic}}\label{app:QB}
In this appendix we summarize the necessary data from Ref.~\cite{quach2020organic} to calculate $\Omega_r$ and $\Delta$ in the non-resonant variational polaron master equation. 

From the `Materials and methods' section of Ref.~\cite{quach2020organic} one finds that the light--matter coupling is $g=10.6~\text{neV}$, $N\in[0.21,16]\times10^{10}$ across the experiments which were performed at room temperature. The dephasing rate is $(1.68/N)~\text{meV}$ which, due to the $1/N$ scaling, means that we must assume an Ohmic spectral density. Using Eq.~\eqref{eq:J} with $p=1$ in Eq.~\eqref{eq:g10} one finds that $A=(4\beta/\pi)\times1.68~\text{meV}=0.083$.

\section{Width of polariton peaks in absorption spectrum}\label{app:spectrum}
In this appendix we derive an expression for the widths of the polariton peaks in terms of quantities discussed in the main text.

The absorption spectrum of the cavity is \cite{ficek2005quantum,nazir2016modelling}
\begin{equation}\label{eq:SpectraCav}
	\ms{A}(\omega)=\text{Re}\inta{\tau}\rme^{i\omega\tau}\lim_{t\to\infty}\langle \mb{E}_+(\mb{R},t+\tau)\cdot\mb{E}_-(\mb{R},t)\rangle,
\end{equation}
which is the Fourier transform of the correlation function between the positive and negative components of the electric field,
\begin{align}
\mb{E}_+(\mb{R},t)=+i\mb{e}\sqrt{\frac{\omega_c}{2V}} a(t)\rme^{i\omega_c R},
\end{align} 
and $\mb{E}_-(\mb{R},t)=\mb{E}_-(\mb{R},t)^\dagger$, where $\mb{R}$ is the position of the detector, $V$ and $\mb{e}$ are the quantization volume and polarization vector of the cavity mode, and $a(t)$ is the photon annihilation operator in the Heisenberg picture.

Assuming that the detector is far enough from the dipole that we can ignore the phase factors, the absorption spectrum is
\begin{equation}\label{eq:abscav}
	\ms{A}(\omega)=\ms{A}_0\text{Re}\inta{\tau}\rme^{i\omega\tau}\lim_{t\to\infty}\langle a(t+\tau) a^\dagger(t)\rangle,
\end{equation}
where $\ms{A}_0=\omega_c/(2V)$. Note that the area of the spectrum is a constant, \begin{equation}\label{eq:Sarea}
\inta{\omega}\ms{A}(\omega)=\pi \ms{A}_0 \left(1+\langle a^\dagger(\infty)a(\infty)\rangle\right).
\end{equation} 

Using the quantum regression theorem \cite{mccutcheon2016optical} in the variational polaron frame, one finds that
\begin{equation}\label{eq:aatrace} 
\lim_{t\to\infty}\langle a(t+\tau)a^\dagger(t)\rangle=\text{Tr}\left[a\zeta(\tau)\right],
\end{equation}
where 
\begin{equation}
	\zeta(\tau)=\lim_{t\to\infty}\text{Tr}_B\left[U_0(\tau)a^\dagger\varrho(t)U_0^\dagger(\tau)\right],
\end{equation} 
and $U_0(\tau)=\exp[-i\ms{H} \tau]$. The operator $\zeta(\tau)$ evolves with respect to $\tau$ under the same master equation as $\varrho(\tau)$---the variational polaron frame master equation---but has the modified initial state,
\begin{equation}\label{eq:zeta0}
	\zeta(0)=a^\dagger\varrho(\infty).
\end{equation}

Within the $n=1$ single exciton manifold the only non-zero term in the trace in Eq.~\eqref{eq:aatrace} comes from $\bra{G,0}a=\bra{G,1}$ such that
\begin{align}
	\lim_{t\to\infty}\langle a(t+\tau)a^\dagger(t)\rangle&=\braket{G,1|\zeta(\tau)|G,0}\\
	&=\frac{1}{\sqrt{2}}\left[\zeta_{+G}(\tau)+\zeta_{-G}(\tau)\right],
\end{align}
where subscript `$G$' refers to the zero excitation state $\ket{G,0}$. The matrix elements $\zeta_{\pm G}(\tau)$ evolve in $\tau$ identically to the time evolution of the coherences between the polariton states $\ket{\pm}$ and the ground state $\ket{G,0}$. Therefore, the absorption spectrum is related to the variational polaron frame master equation by
\begin{equation}\label{eq:SA}
\ms{A}(\omega)=\frac{\ms{A}_0}{\sqrt{2}}\text{Re}\inta{\tau}\rme^{i\omega\tau}\left[\zeta_{+G}(\tau)+\zeta_{-G}(\tau)\right],
\end{equation}
with $\zeta_{\pm G}(0)=\varrho_{GG}(\infty)/\sqrt{2}$ obtained from Eq.~\eqref{eq:zeta0}. Notice that Eq.~\eqref{eq:SA} yields $\inta{\omega}\ms{A}(\omega)=\pi \ms{A}_0 \varrho_{GG}(\infty)$ for the area of the spectrum. When compared to Eq.~\eqref{eq:Sarea} this implies that $\rho_{GG}(\infty)=1$ and $\langle a^\dagger(\infty)a(\infty)\rangle=0$, that is, all photons must eventually leak from the cavity. This appears to contradict our assumption in the main text that the number of photons in the cavity is constant and equal to one. But, we retain consistency if photon leakage from the cavity occurs over a much longer timescale than the vibrational dynamics.

 We must now evaluate the integral in Eq.~\eqref{eq:SA} using the variational polaron frame master equation. To obtain results that we can interpret in terms of parameters in the main text we secularize the variational polaron frame master equation. In a secular master equation, coherences evolve generally as $\dot{\varrho}_{\mu\nu}(\tau)=-R_{\mu\nu}\varrho_{\mu\nu}(\tau)$ with solution
 \begin{equation}\label{eq:VPMEsec}
\varrho_{\mu\nu}(\tau)=\varrho_{\mu\nu}(0)\rme^{-R_{\mu\nu}\tau},
\end{equation}
where $\mu\neq\nu$ and $R_{\mu\nu}$ is given in Eq.~\eqref{eq:Rmunu} for the variational polaron master equation. Compared to Eq.~\eqref{eq:VPME}, we can neglect the other terms within the ellipses in Eq.~\eqref{eq:VPMEsec} because we are only concerned with coherences $\mu\neq\nu$ \textit{and} we have secularized the master equation.

One can now substitute Eq.~\eqref{eq:VPMEsec} into Eq.~\eqref{eq:SA}, replacing $\varrho_{\pm G}(\tau)$ with $\zeta_{\pm G}(\tau)$ and using the initial conditions $\zeta_{\pm G}(0)=1/\sqrt{2}$. Performing the integration yields Eq.~\eqref{eq:A2} in the main text but with the WCME quantities replaced with the VPME quantities: $r_{pG}\to R_{pG}$ and $\delta_{pG}\to\Delta_{pG}$.
The spectrum describes two Lorentzian distributions with maxima at the Lamb shifted polariton energies $\Delta_{\pm G}=\omega_{\pm}+\Lambda_\pm$, and full width half maxima equal to $2\text{Re}[R_{\pm G}]=K^\downarrow_\pm+2K^\phi_\pm$. 

\section{Variational optimisation}\label{app:VO}
In this section we derive the optimisation scheme that determines which variational parameters $\{\eta\uk\}$ give the unperturbed Hamiltonian $\mc{H}_S$ in Eq.~\eqref{eq:HSp} that most closely resembles the equilibrium state of the full Hamiltonian $\mc{H}$. 

The equilibrium density operator of the model is $\exp(-\beta \mc{H})/Z$ where $Z=\text{Tr}_B[\exp(-\beta \mc{H})]$ is the partition function, which has a free energy $F=-\beta^{-1}\ln Z$ that is minimised in equilibrium. Substituting the partition function into $F$ and then using the Feynman-Bogoliubov-Peierls upper bound identity \cite{pollock2013multi} we find that,
\begin{align}\label{eq:F}
F&=-\beta^{-1}\ln \text{Tr}_B\left[\rme^{-\beta\left(\mc{H}_0+\mc{H}_{SB}\right)}\right]\nonumber\\
&\le -\beta^{-1}\ln \text{Tr}_B\left[\rme^{-\beta \mc{H}_0}\right]+\frac{\text{Tr}_B[\mc{H}_{SB}\rme^{-\mc{H}_0}]}{\text{Tr}_B[\rme^{-\mc{H}_0}]}+\mathcal{O}\left(\mc{H}_{SB}^2\right),
\end{align}
where $\mc{H}_0=\mc{H}_S+\mc{H}_B$ is the unperturbed Hamiltonian. By construction, $\text{Tr}_B[\mc{H}_{SB}\exp(-\beta \mc{H}_0)]=0$, and further ignoring terms of second order and greater in the perturbation leads to
\begin{equation}\label{eq:AB}
	F\lesssim-\beta^{-1}\ln\text{Tr}_B\left[\rme^{-\beta \mc{H}_0}\right]\equiv F_\text{FBP}.
	\end{equation}

Eq.~\eqref{eq:AB} indicates that the free energy of system plus bath is less than or equal to the free energy of the unperturbed variational polaron frame Hamiltonian, $F_\text{FBP}$. By choosing $\{\eta\uk\}$ to minimise $F_\text{FBP}$, $\mc{H}_0$ will be defined such that it gives the closest representation of the equilibrium state as permitted by a polaron type transformation. Therefore, the equilibrium contribution of $\mc{H}_{SB}$ to the dynamics will be minimal, and a theory perturbing in $\mc{H}_{SB}$ as accurate as possible through optimising $\{\eta\uk\}$.

Using the eigensystem of $\mc{H}_0$ we find 
\begin{multline}\label{eq:AB2}
F_\text{FBP}=-\frac{1}{\beta}\ln
\bigg(1+\sum_{n=1}^\infty\Big[2\cosh\left[\frac{\beta\theta_n}{2}\right]\rme^{-\beta\left(\frac{\Delta}{2}+n\nu\right)}\\
+ \left(N-1\right)\rme^{-\beta\left(\Delta+n\nu\right)}\Big]\bigg)\text{tr}\left[\rme^{-\beta \mc{H}_B}\right],
\end{multline}
where $\Delta=\omega_m-\lambda-\omega_c$ is the detuning and $\theta_n=\sqrt{\Delta^2+4\Omega^{ 2}_r n}$. Throughout the main text we assume that there is one photon in the cavity which means that only the eigenstates with $n=1$  contribute to the dynamics. The equivalent assumption here is to only take the $n=1$ term in the summation in Eq.~\eqref{eq:AB2}. We then minimise $F_\text{FBP}$ with respect to $\eta\uk$ to find the optimal expression for $G(\omega\uk)=\eta\uk/f\uk$ given by Eqs.~\eqref{eq:OptRes}--\eqref{eq:Gbar} in the main text.

\section{Variational optimization in the low temperature regime}\label{app:LowT}
In the main text we focused on the experimentally relevant, room temperature regime with $\omega_0\lesssim\Omega_\beta$. In this appendix we discuss differences when the temperature is low enough that $\omega_0\gtrsim\Omega_\beta$. For a typical molecular high frequency cutoff of $\omega_0=6~\text{meV}$ \cite{shalabney2015coherent} this requires temperatures below $7~\text{K}$.

\begin{figure}[ht!]\centering
	\includegraphics[width=\columnwidth]{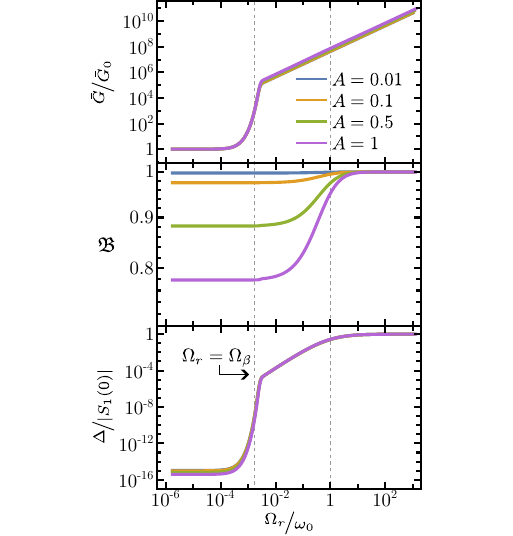}
	\caption{Similar to Fig.~\ref{fig:kappa} but here in the low temperature regime with $\omega_0\gtrsim\Omega_\beta$. Parameters used: $g=0.1~\mu\text{eV}$, $\omega_0=50~\text{meV}$, $T=0.1~\text{K}$, and $p=3$. Note that we use exaggerated values for $\omega_0$ and $T$ to clearly demonstrate the regime $\Omega_\beta<\Omega_r<\omega_0$.} \label{fig:kappa2}
\end{figure}

Fig.~\ref{fig:kappa2} is similar to Fig.~\ref{fig:kappa} in the main text but now within the low temperature regime with $\omega_0\gtrsim\Omega_\beta$. There are two differences between the high and low temperature parameter regimes. (1) At low temperature, Fig.~\ref{fig:kappa2} demonstrates that the effects of increasing vibrational coupling strength are diminished because the baths are essentially `frozen out'. (2) At low temperature, the size of $\Omega_r$ compared to both $\Omega_\beta$ and $\omega_0$ are important for $\mathfrak{B}$ and $\Delta$, whereas at high temperature only $\Omega_\beta$ was important. These differences do not change any of the qualitative conclusions we have drawn in the main text regarding the $N$ scaling of the master equation. This is because in both Fig.~\ref{fig:kappa} and Fig.~\ref{fig:kappa2}, $\bar{G}$ is described by Eq.~\eqref{eq:GN}, and the system is always non-resonant to a good approximation when $\Omega_r\ll\Omega_\beta$.

\section{Master equation derivation}\label{app:ME}
In this appendix we derive the non-secular Redfield equation in the variational polaron frame. The Redfield equation in the Schr\"odinger picture is
\begin{multline}	
	\partial_t\varrho(t)=-i\left[\mc{H}_S,\varrho(t)\right]\\-U_S(t)\int_0^{\infty}\text{d}\tau\ \text{tr}_B\left[\tilde{\mc{H}}_{SB}(t),\left[\tilde{\mc{H}}_{SB}(t-\tau),\varrho(t)\right]\right]U_S(t)^\dagger
\end{multline}
where a tilde denotes operators transformed into the interaction picture, $\text{tr}_B[\cdot]$ is a trace over the Hilbert spaces of the baths, and $U_S(t)=\exp(-i\mc{H}_St)$. $\ms{H}_S$ is the system Hamilton in Eq.~\eqref{eq:HSp} and $\ms{H}_{SB}=\ms{H}_D+\ms{H}_P$ is the system--bath interaction given in Eqs.~\eqref{eq:VE}--\eqref{eq:VP}. For later algebraic ease we decompose the master equation into contributions from each interaction type,
\begin{equation}
	\partial_t\varrho(t)=-i\left[\mc{H}_S,\varrho(t)\right]+\sum_{a,b\in\{D,P\}}L_{ab}[\varrho(t)],
\end{equation}
where
\begin{multline}
	L_{ab}\left[\varrho(t)\right]=\\-U_S(t)\int_0^{\infty}\text{d}\tau\ \text{tr}_B\left[\tilde{\mc{H}}_a(t),\left[\tilde{\mc{H}}_b(t-\tau),\varrho(t)\right]\right]U_S(t)^\dagger.
\end{multline}
The superoperators $L_{DD}[\varrho(t)]$ and $L_{PP}[\varrho(t)]$ are the displacement--type and polaron--type master equations arising solely from $\mc{H}_D$ and $\mc{H}_{P}$, respectively, and $L_{PD}[\varrho(t)]+L_{DP}[\varrho(t)]$ is a superoperator unique to the variational polaron master equation. 

We will now derive each master equation contribution in turn. We use subscripts $\{\alpha,\beta,\gamma,\delta\}$ to denote any eigenstate $\ket{+}$, $\ket{-}$ and $\{\ket{d}\}$ for $d\in\{d_1,\ldots,d_{N-1}\}$ whilst we use $\{p,q,r,s\}$ to label only the polaritons $\ket{+}$ and $\ket{-}$. Finally, we define the transition energies and eigenstate transition operators,
\begin{align}
\omega_{\alpha\beta}&=\omega_\alpha-\omega_\beta,\\
\Pi_{\alpha\beta}&=\proj{\alpha}{\beta}.
\end{align}

\subsection{Displacement--type contribution}
Substituting $\ms{H}_D$ in Eq.~\eqref{eq:VE} into $L_{DD}[\varrho(t)]$ yields the displacement--type master equation which has the same form as the second line of the weak vibrational coupling master equation in Eq.~\eqref{eq:MEweak} of the main text, except that the spectral density in the correlation functions is replaced with the displacement--type spectral density,
\begin{equation}
	J_D(\omega)=J(\omega)\left(1-G(\omega)\right)^2.
\end{equation}
The origin of this difference can be seen by comparing the form of the couplings in the lab and variational frame displacement--type interactions in Eq.~\eqref{eq:HSB} and \eqref{eq:VE}, respectively. That is, one makes the substitution $f\uk\to f\uk-\eta\uk=f\uk(1-G(\omega\uk))$ to move from the weak coupling $H_{SB}$ to the variational frame displacement--type interaction $\ms{H}_D$.

We find that
\begin{equation}\label{eq:MEd}
L_{DD}[\varrho(t)]=	\sum_{\alpha,\beta,\gamma,\delta}c_{\alpha\beta\gamma\delta}\Gamma^D_1(\omega_{\delta\gamma})\left[\Pi_{\gamma\delta}\varrho_S(t),\Pi_{\alpha\beta}\right]+\text{H.c.},
\end{equation}
where $c_{\alpha\beta\gamma\delta}$ is defined in Eq.~\eqref{eq:c} and
\begin{equation}
	\Gamma_1^D(\nu)=M_+\left[J_D(\nu)\right],
\end{equation}
where 
\begin{equation}\label{eq:RM}
\text{Re}\left(M_\pm\left[F(\nu)\right]\right)=\pi \times\begin{cases}
	\pm F(\nu)\tilde{n}_B(\nu)&\text{if }\nu\ge0,\\
	F(-\nu)n_B(-\nu)&\text{if }\omega<0,
	\end{cases}
\end{equation}
and
\begin{multline}\label{eq:IM}
\text{Im}\left(M_\pm\left[F(\nu)\right]\right)=\\
\mathcal{P}\inta{\omega}F(\omega)\left[\pm\frac{\tilde{n}_B(\omega)}{\nu-\omega}+\frac{n_B(\omega)}{\nu+\omega}\right],
\end{multline}
for any function $F(\nu)$. The functional with a negative subscript, $M_-[F(\nu)]$, will be used in the variational--type master equation. In this notation, the Fourier transform of the single phonon correlation function given in Eq.~\eqref{eq:G1} is $\Gamma_1(\nu)=M_+[J(\nu)]$.

\subsection{Polaron--type contribution}
Substituting $\ms{H}_{P}$ in Eq.~\eqref{eq:VP} into $L_{PP}[\varrho(t)]$ yields,
\begin{align}\label{eq:MEp}
	L_{PP}&\left[\varrho(t)\right]=\\-\sum_{\alpha\beta pq}
	&\Big(c^{P(-)}_{\alpha\beta}\Gamma^{P(-)}(\omega_{q\beta})\left[\Pi_{\alpha p},\Pi_{\beta q}\varrho_S(t)\right]\\
	&+c^{P(-)*}_{\alpha\beta}\Gamma^{P(-)}(\omega_{\beta q})\left[\Pi_{p\alpha},\Pi_{q\beta}\varrho_S(t)\right]\\
&+c^{P(+)}_{\alpha\beta}\Gamma^{P(+)}(\omega_{\beta q})\left[\Pi_{\alpha p},\Pi_{q\beta}\varrho_S(t)\right]\\
&+c_{\alpha\beta}^{P(+)*}\Gamma^{P(+)}(\omega_{q\beta})\left[\Pi_{p\alpha},\Pi_{\beta q}\varrho_S(t)\right]\Big)+\text{H.c.},	
	\end{align}
where the rate functions are
\begin{equation}\label{eq:GP}
\Gamma^{P(\pm)}(\nu)=\frac{\Omega_r^2}{2N}\inta{\tau}\rme^{i\nu\tau}\left(\rme^{\pm\phi(\tau)}-1\right),
\end{equation}
and the coefficients are,
\begin{align}
	c^{P(-)}_{\alpha\beta}&=\sum_{i=1}^Nu_{i\alpha}u_{i\beta},\label{eq:c-}\\
	c^{P(+)}_{\alpha\beta}&=\sum_{i=1}^Nu_{i\alpha}u_{i\beta}^*.\label{eq:c+}
\end{align}
The polaron--type phonon propagator $\phi(\tau)$ is defined in Eq.~\eqref{eq:phi} and depends on the polaron--type spectral density function,
\begin{equation}\label{eq:JP}
	J_P(\omega)=J(\omega)\frac{G(\omega)^2}{\omega^2}.
\end{equation}

By expanding $\exp(\phi(\tau))\approx 1+\phi(\tau)$ in Eq.~\eqref{eq:GP}, one can identify the single and multi phonon contributions of the polaron--type master, by using
\begin{multline}
    \inta{\tau}\rme^{i\nu\tau}\left(\rme^{\pm\phi(\tau)}-1\right)=\pm M_+\left[J_P(\nu)\right]\\+\inta{\tau}\rme^{i\nu\tau}\left(\rme^{\pm\phi(\tau)}-1\mp\phi(\tau)\right).
\end{multline}

\subsection{Variational--type contribution}
Substituting $\ms{H}_D$ and $\ms{H}_P$ in Eqs.~\eqref{eq:VE} and \eqref{eq:VP} into $L_{DP}[\varrho(t)]+L_{PD}[\varrho(t)]$ yields
\begin{align}\label{eq:MEv}
	&L_{DP}[\varrho(t)]+L_{PD}[\varrho(t)]=\\&-\sum_{\alpha\beta\gamma p}\Big(c^{V}_{1,\alpha\beta\gamma}\Gamma^V(\omega_{\gamma p})\left[\Pi_{\alpha\beta},\Pi_{p\gamma}\varrho_S(t)\right]\\
	&\qquad+c^{V}_{2,\alpha\beta\gamma}\Gamma^V(\omega_{\beta\alpha})\left[\Pi_{\gamma p},\Pi_{\alpha\beta}\varrho_S(t)\right]\\
	&\qquad+c_{1,\alpha\beta\gamma}^{V}\Gamma^V(\omega_{\beta\alpha})\left[\Pi_{\alpha\beta}\varrho_S(t),\Pi_{p\gamma}\right]\\
	&\qquad+c_{2,\alpha\beta\gamma}^{V}\Gamma^V(\omega_{p\gamma})\left[\Pi_{\gamma p}\varrho_S(t),\Pi_{\alpha\beta}\right]\Big)+\text{H.c.},
\end{align}	 
where the rate function is
\begin{equation}
	\Gamma^V_{\alpha\beta\gamma}(\nu)=\frac{\Omega_r}{\sqrt{2N}}M_-\left[J_V(\nu)\right],
\end{equation}
the coefficients are,
\begin{align}
c_{1,\alpha\beta\gamma}^{V}&=\sum_{i=1}^Nu_{i\alpha}u_{i\beta}^*u_{i\gamma}^*\label{eq:cV1}\\
c_{2,\alpha\beta\gamma}^{V}&=\sum_{i=1}^Nu_{i\alpha}u_{i\beta}^*u_{i\gamma}\label{eq:cV2},
\end{align}
and the real and imaginary parts of $M_-[\cdot]$ are given in Eqs.~\eqref{eq:RM}--\eqref{eq:IM}. The variational--type spectral density function is
\begin{align}
J_V(\omega)&=J(\omega)\left(1-G(\omega)\right)\frac{G(\omega)}{\omega}\\
&=\sqrt{J_D(\omega)J_P(\omega)}.
\end{align}

\section{Non resonance}\label{app:resonance}
As shown in Fig.~\ref{fig:kappa} in the main text, if $\Omega_r \gg\Omega_\beta$ then the detuning becomes equal to the vibrational reorganization energy, $\Delta=-S_1(0)$. If the vibrational coupling is strong enough that $|\Delta|$ is comparable to $2\Omega_r$, then the system must be described by a non-resonant Hamiltonian in the variational polaron frame.

If one cannot make the resonant approximation, then within the single photon and exciton manifold, $\mc{H}_S$ in Eq.~\eqref{eq:HSp} has the following polariton eigenstates,
\begin{align}
	\ket{\pm}=\mp\sqrt{N}U_\mp\ket{G,1}\pm U_\pm\ket{B},
\end{align}
where
\begin{equation}\label{eq:upm}
	U_\pm=\frac{\pm 1}{\sqrt{2N}}\left(1\pm\frac{\Delta}{\theta}\right)^\frac{1}{2},
\end{equation}
and 
\begin{equation}
	\theta=\sqrt{4\Omega_r^2+\Delta^2}.
\end{equation}
The polariton states have energies
\begin{equation}
	\omega_\pm=\frac{\omega_m+\omega_c\pm\theta}{2}.
\end{equation}
The $N-1$ degenerate dark states are described by the same vectors as in the resonant model and have an energy $\omega_m$. This means that the transition energies in the non-resonant model are asymmetric, $\omega_+-\omega_d=(\theta-\Delta)/2$ and $\omega_d-\omega_-=(\theta+\Delta)/2$.

We will now discuss the transition rates, dephasing rates, and Lamb shifts for the non-resonant variational polaron master equation. As in the resonant case in the main text, we derive these quantities by deriving the following element of the master equation
\begin{equation}
	\dot{\varrho}_{\mu\nu}(t)= -R_{\mu\nu}(\Delta) \varrho_{\mu\nu}(t)+\ldots,
\end{equation}
where 
\begin{equation}\
	R_{\mu\nu}(\Delta)=\frac{K_\mu^\downarrow(\Delta)+K_{\nu}^\downarrow(\Delta)}{2}+K^\phi_{\mu\nu}(\Delta)+i\Delta_{\mu\nu}(\Delta).
\end{equation}
$R_{\mu\nu}(0)=R_{\mu\nu}$ is the resonant value given in Section~\ref{sec:VPT}. The loss rates can be written as summations of the transition rates,
\begin{equation}
	K_\mu^\downarrow(\Delta)=\sum_{\alpha\neq\mu}K_{\mu\to\alpha}(\Delta),
\end{equation}
and the Lamb shifted transition frequencies are
\begin{equation}
	\Delta_{\mu\nu}(\Delta)=\left[\omega_\mu(\Delta)+\Lambda_\mu(\Delta)\right]-\left[\omega_\nu(\Delta)+\Lambda_\nu(\Delta)\right],
\end{equation}
where $\omega_\mu(\Delta)$ are the non-resonant eigenenergies.

We can anticipate the effects of detuning by considering the coupling operators in $\ms{H}_{SB}=\ms{H}_D+\ms{H}_P$ in Eqs.~\eqref{eq:VE} and \eqref{eq:VP}. In the large detuning limit, the lower polariton localizes onto the single photon state $\ket{-}\to\ket{G,1}$, whilst the upper polariton localizes onto the bright state $\ket{+}\to\ket{B}$. Therefore, in $\ms{H}_D$---which describes single phonon processes---the molecular coupling operator $\sigma^+_i\sigma^-_i=\proj{e_i,0}{e_i,0}$ only connects the dark states and upper polariton together, whilst in $\ms{H}_P$---which describes single and multi phonon processes---the coupling operator $a\sigma_i^+=\proj{e_i,0}{G,1}$ only connects the dark states and upper polariton to the lower polariton. Consequently, in the large detuning limit, we expect single phonon processes involving the lower polariton to be suppressed, and multi phonon processes between the upper polariton and dark states to be suppressed. Additionally, since $a\sigma_i^+$ does not contain a state projector in the large detuning limit, we also expect multi phonon dephasing processes to be suppressed. As discussed in the main text, the leading order contribution to dephasing is multi phonon, and so non-resonance will lead to narrower polariton line widths than expected from the resonant theory.

\subsection{Master equations}
The master equations for the non-resonant model are modified slightly from those given in Appendix~\ref{app:ME} for the resonant model, due to the modified eigenstates.

The displacement--type master equation has the same form as Eq.~\eqref{eq:MEd} with $c_{\alpha\beta\gamma\delta}$ given by Eq.~\eqref{eq:c}. However, the $u_{i\alpha}$ now take the forms
\begin{equation}\label{eq:unres}
    u_{i\alpha}=\begin{cases}
        U_\pm&\text{ if }\alpha=\pm,\\
        u_{id}&\text{ if }\alpha=d,
    \end{cases}
\end{equation}
where $U_\pm$ are given in Eq.~\eqref{eq:upm} and $u_{id}$ are the same as in the resonant theory. 

The non-resonant polaron--type master equation can be obtained from Eq.~\eqref{eq:MEp} with the replacements,
\begin{equation}
    c_{\alpha\beta}^{P(\pm)}\to c_{\alpha\beta pq}^{P(\pm)}=2Nc_{\alpha\beta}^{P(\pm)}|U_{-p}||U_{-q}|,
\end{equation}
where $c_{\alpha\beta}^{P(\pm)}$ are given in Eqs.~\eqref{eq:c-}--\eqref{eq:c+}. 

Lastly, the non-resonant variational--type master equation can be obtained from Eq.~\eqref{eq:MEv} with the replacements,
\begin{equation}
    c_{j,\alpha\beta\gamma}^{V}\to c_{j,\alpha\beta\gamma p}^{V}=\sqrt{2N}c_{j,\alpha\beta\gamma}^{V}|U_{-p}|,
\end{equation}
for $j\in\{1,2\}$ where $c_{j,\alpha\beta\gamma}^{V}$ are given in Eqs.~\eqref{eq:cV1}--\eqref{eq:cV2}.

\subsection{Transition rates}
We will write the rates in terms of a generalized rate function,
\begin{align}\label{eq:G}
	\gamma_\Delta(\nu,\{a,b,c\})=a\ \gamma_1(\nu)+b\ \gamma_{>1}^\text{even}(\nu)+c\ \gamma_{>1}^\text{odd}(\nu)
\end{align}
where $a$, $b$, and $c$ are free parameters that may depend on $\Delta$. $\gamma_1(\nu)$ describes single phonon processes and is given in Eq.~\eqref{eq:g1}. The remaining functions $\gamma_{>1}^\text{even}(\nu)=2\text{Re}[\Gamma_{>1}^\text{even}(\nu)]$ and $\Gamma_{>1}^\text{odd}(\nu)=2\text{Re}[\Gamma_{>1}^\text{odd}(\nu)]$ describe even- and odd-ordered multi phonon processes, given generally by
\begin{align}\label{eq:Gg1M}
	\Gamma_{>1}^M(\nu)=\Omega_r^2\sum_{\substack{m=2\\m\in M}} \frac{1}{m!}\inta{\tau}\rme^{i\nu\tau}\phi(\tau)^m,
\end{align}
where $M\in\{\text{even},\text{ odd}\}$ denotes only even or odd values of $m$ are included in the summation, and $m=2$ is excluded if $M\in\text{odd}$. We also define the dimensionless parameter,
\begin{equation}
	\varepsilon=\frac{\Delta}{\theta},
\end{equation}
which quantifies the detuning.

We find the non-resonant polariton-to-polariton transition rates,
\begin{equation}
	K_{\pm\to\mp}(\Delta)=\frac{1}{4N}\gamma_\Delta\left(\pm \theta,\left\{1-\varepsilon^2,2\varepsilon^2,2\right\}\right),\label{eq:Kpmpm}
\end{equation} 
which indicates that in the limit of large detuning, single phonon transitions between polaritons are suppressed, whilst even ordered multi phonon transitions are enhanced. The transition rates from the polaritons to the dark states are
\begin{multline}\label{eq:KpmdDelta}
	K_{\pm\to d}(\Delta)=\\\frac{1}{2N}\gamma_\Delta\left(\pm\frac{\theta\mp\Delta}{2},\left\{1\pm\varepsilon,1\mp\varepsilon,1\mp\varepsilon\right\}\right),
\end{multline}
and transitions from dark states to the polaritons are the same up to a sign change on the first argument of the function on the right-hand-side which turns absorption processes into emission and vice-versa. The transition rates between degenerate dark states are the same as in the resonant model because dark states do not change off-resonance, $K_{d\to d'\neq d}(\Delta)=\gamma_1(0)/N$. 

Eqs.~\eqref{eq:Kpmpm}--\eqref{eq:KpmdDelta} show that as the detuning increases, transitions between the upper polariton and the dark states are increasingly dominated by single phonon processes, whilst multi phonon processes become increasingly dominant for transitions between the lower polariton and the dark states. In the limit $\Omega_r\gg\omega_0$ this may have important implications because multi phonon processes are exponentially faster than single phonon processes. The total loss rates from the eigenstates are,
\begin{align}
	K_\pm^\downarrow(\Delta)&=K_{\pm\to \mp}(\Delta)+(N-1)K_{\pm\to d}(\Delta),\\
	K_d^\downarrow(\Delta)&=K_{d\to +}(\Delta)+K_{d\to -}(\Delta)+\frac{N-2}{N}\gamma_1(0).
\end{align}

\subsection{Dephasing rates}
As in the resonant master equation, the dephasing rates have contributions from the displacement--type and polaron--type master equations:
\begin{equation}
	K_{\mu\nu}^\phi(\Delta)=k_{\mu\nu}^\phi(\Delta)+k_{\mu\nu}^{\Phi}(\Delta).
\end{equation}
The displacement--type contribution has the same form as in Eq.~\eqref{eq:gWPhi} but with the $c_{\alpha\beta\gamma\delta}$ coefficient in Eq.~\eqref{eq:c} now dependent on the non-resonant eigenbasis as described by Eq.~\eqref{eq:unres}. The polaron--type contribution---with $k_{\mu\nu}^\Phi(0)$ given for the resonant model in Eq.~\eqref{eq:gPphi}---gains an overall prefactor dependent on the detuning,
\begin{equation}
	k_{\mu\nu}^{\Phi}(\Delta)=\left(1-\varepsilon^2\right)k_{\mu\nu}^{\Phi}(0).
\end{equation}
In terms of the generalized dephasing function,
\begin{equation}
	\gamma_\Delta^\phi\left(\{a,b\}\right)=a\ \gamma_1(0)+b\ \gamma_{>1}^\phi(0),
\end{equation}
where $\gamma_{>1}^\phi(0)$ is the multi phonon dephasing rate function defined through Eq.~\eqref{eq:gP0}, the dephasing rates are
\begin{align}
	K_{+-}^\phi(\Delta)&=\gamma^{\phi}_\Delta\left(\left\{\frac{1}{2N}\varepsilon^2,2\left(1-\varepsilon^2\right)\right\}\right),\\
	K_{\pm G}^\phi(\Delta)&=\gamma^\phi_\Delta\left(\left\{\frac{1}{8N}\left(1\pm\varepsilon\right)^2,\frac{1}{2}\left(1-\varepsilon^2\right)\right\}\right),\label{eq:KpmPhiDelta}\\
	K_{\pm d}^\phi(\Delta)&=\gamma^\phi_\Delta\left(\left\{\frac{1}{8N}\left(1\mp\varepsilon\right)^2,\frac{1}{2}\left(1-\varepsilon^2\right)\right\}\right),\\
	K_{dG}^\phi(\Delta)&=K_{dG}^\phi(0)=\gamma_\Delta^\phi\left(\left\{\frac{1}{2N},0\right\}\right),\\
	K_{d_id_j}^\phi(\Delta)&=K_{d_id_j}^\phi(0)=0.
\end{align}
When the detuning is large, multi phonon dephasing processes are suppressed by $1-\varepsilon^2$. Moreover, Eq.~\eqref{eq:KpmPhiDelta} shows that non-resonance breaks the equality of the polariton dephasing rates such that $K_{+G}^\phi(\Delta)> K_{-G}^\phi(\Delta)$. However, the symmetry breaking occurs in the single phonon dephasing processes, which are not the leading order contribution, and so this effect may be too small to observe even for a large detuning.

\subsection{Lamb shifts}
We will write the Lamb shifts in terms of a generalized Lamb shift function,
\begin{equation}\label{eq:SDel}	
	S_\Delta(\nu,\{a,b,c\})=a\ S_1^v(\nu)+b\ S_{>1}^\text{even}(\nu)+c\ S_{>1}^\text{odd}(\nu)
\end{equation}
where $S_1^v(\nu)$ describes single phonon processes and is given in Eq.~\eqref{eq:S1v}. The remaining functions are $S_{>1}^\text{even}(\nu)=\text{Im}[\Gamma_{>1}^\text{even}(\nu)]$ and $S_{>1}^\text{odd}(\nu)=\text{Im}[\Gamma_{>1}^\text{odd}(\nu)]$ where $\Gamma_{>1}^\text{even}(\nu)$ and $\Gamma_{>1}^\text{odd}(\nu)$ are given in Eq.~\eqref{eq:Gg1M}. In terms of Eq.~\eqref{eq:SDel} the Lamb shift induced by transitions between polaritons is
\begin{equation}
	\Lambda^t_{\pm\to\mp}(\Delta)=\frac{1}{4N}S_\Delta(\pm\theta,\{1-\varepsilon^2,2\varepsilon^2,2\}),
\end{equation}
and the shift induced by transitions from polaritons to dark states is
\begin{equation}\label{eq:LdpmNR}
	\Lambda^t_{\pm\to d}(\Delta)=\frac{1}{2N}S_\Delta\left(\pm\frac{\theta\mp\Delta}{2},\{1\pm\varepsilon,1\mp\varepsilon,1\mp\varepsilon\}\right).
\end{equation}
The superscript `$t$' denotes that this arises from real transitions. The Lamb shifts induced by transitions from dark states to polaritons are obtained from Eq.~\eqref{eq:LdpmNR} by inverting the sign in the first argument of the function on the right-hand-side. The Lamb shift induced by transitions between the degenerate dark states is equal to the resonant value, $\Lambda^t_{d\to d'\neq d}=S_1(0)/N$. Notice that Lamb shifts induced by transitions have the same dependencies on detuning as the transitions which generate them, described in Eqs.~\eqref{eq:Kpmpm} and \eqref{eq:KpmdDelta}. 

Lastly, we describe the Lamb shifts induced by virtual self transitions through the generalized function
\begin{equation}
	S_\Delta^\phi(\{a,b\})=a\ S_1^v(0)+b\ S_{>1}^\phi(0),
\end{equation}
where $S_{>1}^\phi(0)$ is the multi phonon contribution defined through Eq.~\eqref{eq:gP0}. One finds the following virtual self Lamb shifts,
\begin{align}
	\Lambda^s_{\pm\to\pm}(\Delta)&=\frac{1}{4N}S_\Delta^\phi\left(\{1\pm\epsilon,1-\epsilon^2\}\right),\\
	\Lambda^s_{d\to d}(\Delta)&=\Lambda^s_{d\to d}(0)=\frac{1}{N}S_\Delta^\phi\left(\{1,0\}\right).\\
\end{align}

Combining the Lamb shifts from virtual and real transitions, the total shift of each eigenstate is
\begin{align}
	\Lambda_\pm(\Delta)&=\Lambda_{\pm\to\mp}(\Delta)+(N-1)\Lambda_{\pm\to d}(\Delta)\nonumber\\
 &\hspace{4.3cm}+\Lambda_{\pm\to\pm}(\Delta),\\
	\Lambda_d(\Delta)&=\Lambda_{d\to+}(\Delta)+\Lambda_{d\to -}(\Delta)+\frac{N-2}{N}S_1(0)\nonumber\\
	&\hspace{4.3cm}+\Lambda_{d\to d}(\Delta),
\end{align}
where each Lamb shift is a summation of the contribution from real and virtual transitions, e.g., $\Lambda_{+\to-}(\Delta)=\Lambda^t_{+\to-}(\Delta)+\Lambda_{+\to-}^s(\Delta)$.

\end{document}